\documentclass[preprint,prd,aps,showpacs,showkeys,nofootinbib]{revtex4}
\usepackage{graphicx}
\usepackage{dcolumn}
\usepackage{bm}
\begin{document}


\title{Electroweak and supersymmetric two-loop corrections to lepton
anomalous magnetic and electric dipole moments}

\author{Tai-Fu Feng\footnote{email: fengtf@dlut.edu.cn}, Lin Sun, Xiu-Yi Yang}

\affiliation{Department of Physics, Dalian University of Technology,
Dalian, 116024, China}

\date{\today}

\begin{abstract}
Using the effective Lagrangian method, we analyze the electroweak
corrections to the anomalous dipole moments of lepton from some
special two-loop diagrams where a closed neutralino/chargino loop is
inserted into relevant two Higgs doublet one-loop diagrams in the minimal
supersymmetric extension of the standard model with CP violation.
Considering the translational invariance of loop momenta and the
electromagnetic gauge invariance, we get all dimension 6 operators and derive
their coefficients. After applying equations of motion
to the external leptons, we obtain the anomalous dipole moments of lepton.
The numerical results imply that there is parameter space where the contributions to the muon
anomalous dipole moments from this sector may be significant.
\end{abstract}


\maketitle

\section{Introduction \label{sec1}}
\indent\indent
At both aspects of experiment and theory, the magnetic dipole moments (MDMs) of lepton
draw the great attention of physicists because of their obvious
importance. The anomalous dipole moments of lepton not only can be
used for testing loop effect in the standard model (SM), but also
provide a potential window to detect new physics beyond the SM.
The current experimental result of the muon MDM is \cite{exp,Rafael1}
\begin{eqnarray}
&&a_{_\mu}^{exp}=11\;659\;208\;\pm\;6\;\times 10^{-10}\;.
\label{data}
\end{eqnarray}

From the theoretical point of view, contributions to the muon MDM are generally
divided into three sectors \cite{Rafael1,Jegerlehner}: QED loops, hadronic contributions
and electroweak  corrections. The largest uncertainty of the SM prediction
originates from the evaluation of hadronic vacuum polarization and
light-by-light corrections. Depending on which evaluation
of hadronic vacuum polarization is chosen, the differences between the SM
predictions and experimental result are given as \cite{Rafael1,Jegerlehner}:
\begin{eqnarray}
&&a_{_\mu}^{exp}-a_{_\mu}^{SM}=33.2\;\pm\;8.8\;\times 10^{-10}\;:\;\;3.8\sigma,
\nonumber\\
&&a_{_\mu}^{exp}-a_{_\mu}^{SM}=30.5\;\pm\;9.3\;\times 10^{-10}\;:\;\;3.3\sigma,
\nonumber\\
&&a_{_\mu}^{exp}-a_{_\mu}^{SM}=28.2\;\pm\;8.9\;\times 10^{-10}\;:\;\;3.2\sigma,
\nonumber\\
&&a_{_\mu}^{exp}-a_{_\mu}^{SM}=11.9\;\pm\;9.5\;\times 10^{-10}\;:\;\;1.3\sigma\;.
\label{sm}
\end{eqnarray}
For the convenience of numerical discussion, we will adopt the second
value in Eq.\ref{sm}. Within three standard error deviations, this difference
implies that the present experimental data can tolerate new physics
correction to the muon MDM as
\begin{eqnarray}
&&2.6\times 10^{-10}\leq\Delta a_{_\mu}^{NP}\leq 58.4\times 10^{-10}\;.
\label{toleration}
\end{eqnarray}

In fact, the current experimental precision ($6\times 10^{-10}$)
already puts very restrictive bounds on new physics scenarios.
In the SM, the electroweak one- and two-loop contributions amount to $19.5\times
10^{-10}$ and $-4.4\times10^{-10}$ respectively.
Comparing with the standard electroweak corrections, the electroweak
corrections  from new physics are generally suppressed by $\Lambda_{_{\rm EW}}^2/\Lambda^2$,
where $\Lambda_{_{\rm EW}}$ denotes the electroweak energy scale and
$\Lambda$ denotes the energy scale of new physics.

Supersymmetry (SUSY) has been considered as a most prospective candidate for
new physics beyond the SM. In the minimal supersymmetric extension of the SM (MSSM)
with CP conservation, the supersymmetric
one-loop contribution is approximately given by
\begin{eqnarray}
\Delta a_{_\mu}^{1L}\simeq13\times10^{-10} \bigg({100\;{\rm
GeV}\over\Lambda}\bigg)^2\tan\beta{\rm sign}(\mu_{_H}),
\label{1L-approxi}
\end{eqnarray}
when all supersymmetric masses are assumed to equal a common mass
$\Lambda$, and $\tan\beta=\upsilon_2/\upsilon_1\gg1$.
Where $\upsilon_1$ and $\upsilon_2$ are the absolute values of the vacuum expectation values
(VEVs) of the Higgs doublets and $\mu_{_H}$ denotes the $\mu$-parameter in
the superpotential of MSSM. It is obvious that
the supersymmetric effects can easily account for the deviation
between the SM prediction and the experimental data.

Actually, the two-loop electroweak corrections to the anomalous
dipole moments of lepton are discussed extensively in literature.
Utilizing the heavy mass expansion approximation (HME) together
with the corresponding projection operator method, Ref.\cite{czarnecki}
has obtained the two-loop standard electroweak correction to the muon MDM
which eliminates some of the large logarithms that were incorrectly kept in
a previous calculation \cite{KKSS}.
Within the framework of MSSM with CP conservation, the authors of Ref. \cite{heinemeyer1,heinemeyer2}
present the supersymmetric corrections from some special two-loop diagrams
where a close chargino (neutralino) loop or a scalar fermion loop is
inserted into those two-Higgs-doublet one-loop diagrams. Ref. \cite{geng}
discusses the contributions to the muon MDM from the effective
vertices $H^\pm W^\mp\gamma, h_0(H_0)\gamma\gamma$ which are induced
by the scalar quarks of the third generation in the CP conserving MSSM.

\begin{figure}[t]
\setlength{\unitlength}{1mm}
\begin{center}
\begin{picture}(0,120)(0,0)
\put(-60,-40){\includegraphics{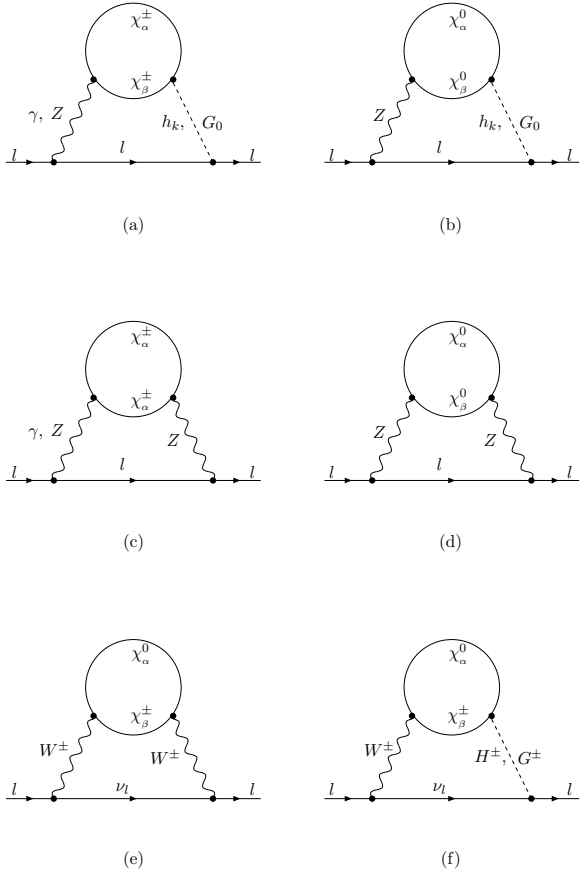}}
\end{picture}
\caption[]{Some two-loop self energy diagrams which lead to the
lepton MDMs and EDMs in  CP violating MSSM, the corresponding triangle diagrams
are obtained by attaching a photon in all possible ways
to the internal particles. In concrete calculation, the contributions from
those mirror diagrams should be included also.}
\label{fig1}
\end{center}
\end{figure}

In this paper, we investigate the electroweak
corrections to the anomalous dipole moments of lepton from some
special two-loop diagrams where a closed neutralino/chargino loop is
inserted into relevant two Higgs doublet one-loop diagrams in the CP violating
MSSM (Fig.\ref{fig1}). Since the masses of those virtual fields
($W^\pm,\;Z$ gauge bosons, neutral and charged Higgs, as well as neutralinos
and charginos) are much heavier than the muon mass $m_\mu$,
we can apply the effective Lagrangian method to get the
anomalous dipole moments of lepton. After integrating out the heavy freedoms mentioned above
and then matching between the effective theory and the full theory, we derive the relevant
higher dimension operators as well as the corresponding Wilson coefficients.
The effective Lagrangian method
has been adopted to calculate the two-loop supersymmetric corrections
to the branching ratio of $b\rightarrow s\gamma$ \cite{Feng1}, neutron EDM \cite{Feng2}
and lepton MDMs and EDMs \cite{Feng3}. In concrete
calculation, we assume that all external leptons as well as photon
are off-shell, then expand the amplitude of corresponding triangle
diagrams according to the external momenta of leptons and photon.
Using loop momentum translational invariance, we formulate the sum of
amplitude from those triangle diagrams which correspond to the corresponding self-energy in
the form which explicitly satisfies the Ward identity required by
the QED gauge symmetry. Then we can get all dimension 6 operators
together with their coefficients. After the equations of
motion are applied to external leptons, higher dimensional operators, such
as dimension 8 operators, also contribute to the muon MDM  and
the electron EDM in principle. However, the contributions of dimension 8
operators contain an additional suppression factor
$m_l^2/\Lambda^2$ comparing with that of dimension
6 operators, where $m_l$ is the mass of lepton. Setting $\Lambda\sim100{\rm GeV}$,
one obtains easily that this suppression factor is
about $10^{-6}$ for the muon lepton. Under current experimental precision, it implies
that the contributions of all higher dimension operators ($D\ge8$)
can be neglected safely.

We adopt the naive dimensional regularization with the
anticommuting $\gamma_{_5}$ scheme, where there is no distinction
between the first 4 dimensions and the remaining $D-4$ dimensions.
Since the bare effective Lagrangian contains the ultraviolet
divergence which is induced by divergent subdiagrams, we give the
renormalized results in the on-mass-shell scheme \cite{onshell}.
Additional, we adopt the nonlinear $R_\xi$ gauge with $\xi=1$ for
simplification \cite{nonlinear-R-xi}. This special gauge-fixing term
guarantees explicit electromagnetic gauge invariance throughout the calculation,
not just at the end because the choice of gauge-fixing term
eliminates the $\gamma W^\pm G^\mp$ vertex in the Lagrangian.

Within the framework of CP violating MSSM, the renormalization-group
improved loop effects of soft CP violating Yukawa interactions related to
scalar quarks of the third generation cause the strong mixing among
CP-even and CP-odd neutral Higgs. The linear expansions of the Higgs
doublet $H^1$ and $H^2$ around the ground state are generally written as
\begin{eqnarray}
&&H^1=\left(\begin{array}{c}{1\over\sqrt{2}}(\upsilon_1+\phi_1^0+ia_1)\\
\phi_1^-\end{array}\right)\;,\;\;
H^2=e^{i\theta}\left(\begin{array}{c}\phi_2^+\\{1\over\sqrt{2}}(\upsilon_2+\phi_2^0+ia_2)
\end{array}\right)\;,
\label{Higgs-doublet}
\end{eqnarray}
where $\theta$ is their relative phase.
In the weak basis $\{\phi_1^0,\;\phi_2^0,\;a=\sin\beta a_1+\cos\beta a_2\}$,
the neutral mass-squared matrix $M_{_H}^2$ may be expressed as
\begin{eqnarray}
&&M_{_H}^2=\left(\begin{array}{ccc}({\cal M}_S^2)_{11}&({\cal M}_S^2)_{12}
&{1\over\cos\beta}({\cal M}_{SP}^2)_{12}\\
({\cal M}_S^2)_{12}&({\cal M}_S^2)_{22}&-{1\over\sin\beta}({\cal M}_{SP}^2)_{21}\\
{1\over\cos\beta}({\cal M}_{SP}^2)_{12}&-{1\over\sin\beta}({\cal M}_{SP}^2)_{21}
&-{1\over\sin\beta\cos\beta}({\cal M}_{SP}^2)_{12}\end{array}\right).
\label{neutral mass-squared matrix}
\end{eqnarray}
Here, the concrete expressions of $({\cal M}_S^2)_{ij},\;({\cal M}_{SP}^2)_{ij}$
can be found in the literature \cite{Pilaftsis}. Since the Higgs mass matrix
$M_H^2$ is symmetric, we can diagonalize it by an orthogonal rotation
$Z_{_H}$ as:
\begin{eqnarray}
&&Z_{_H}^TM_{_H}^2Z_{_H}={\rm diag}(m_{_{h_1}}^2,\;m_{_{h_2}}^2,\;m_{_{h_3}}^2).
\label{rotation matrix}
\end{eqnarray}
Because of this strong mixing among the neutral Higgs, the couplings involving
neutral Higgs are modified drastically comparing with that in CP conservating MSSM.
Certainly, some diagrams in Fig.\ref{fig1} have been discussed in Ref.\cite{heinemeyer2}
where the authors apply the projecting operators to get the lepton MDMs
(Eq.8$\sim$Eq.10 in Ref.\cite{heinemeyer2}) within the framework of CP conservating
MSSM. On the other hand, the fermion electric dipole moments (EDMs) also offer a powerful probe
for new physics beyond the Standard Model (SM). In the SM, the EDMs
of leptons are fully induced by the CP phase of the
Cabibbo-Kobayashi-Maskawa (CKM) matrix elements and they are predicted to be much
smaller \cite{smt} than the present experimental precision \cite{exp1,nexp} and beyond the reach of
experiments in the near future. As for the MSSM, there are many new
sources of the CP violation that can result in larger
contributions to the EDMs of electron and neutron \cite{cp1,cp2}. Taking the
CP phases with a natural size of ${\cal O}(1)$, and the
supersymmetry mass spectra at the TeV range, we can find that the theoretical
predictions on the electron and neutron EDMs at one-loop level already exceed
the present experimental upper bound. In order to make the
theoretical prediction consistent with the experimental data, one can generally
adopt three approaches. One possibility is
to make the CP phases sufficiently small, i.e. $\le 10^{-2}$
\cite{cp1}. One can also assume a mass
suppression by making the supersymmetry spectra heavy, i.e. in the
several TeV range \cite{cp2}, or invoke a cancellation
among the different contributions to the fermion EDMs \cite{cp3}.
Since the lepton EDM is an interesting topic in both theoretical and experimental aspects,
we as well present the lepton EDM by keeping all possible CP violating phases.

This paper is composed by the sections as follows.
In section \ref{sec2}, we introduce the effective Lagrangian method and our notations.
Then we will demonstrate how to obtain the supersymmetric two-loop corrections
to the lepton MDMs and EDMs.
Section \ref{sec3} is devoted to the numerical analysis and
discussion. In section \ref{sec4}, we give our conclusion. Some
tedious formulae are collected in appendix.

\section{Notations and two-loop supersymmetric corrections \label{sec2}}
\indent\indent
The lepton MDMs and EDMs can actually be expressed as the operators
\begin{eqnarray}
&&{\cal L}_{_{MDM}}={e\over4m_{_l}}\;a_{_l}\;\bar{l}\sigma^{\mu\nu}
l\;F_{_{\mu\nu}}
\;,\nonumber\\
&&{\cal L}_{_{EDM}}=-{i\over2}\;d_{_l}\;\bar{l}\sigma^{\mu\nu}\gamma_5
l\;F_{_{\mu\nu}}\;.
\label{adm}
\end{eqnarray}
Here, $\sigma_{\mu\nu}=i[\gamma_\mu,\gamma_\nu]/2$, $l$ denotes the lepton
fermion, $F_{_{\mu\nu}}$ is the electromagnetic field strength, $m_{_l}$ is
the lepton mass and $e$ represents the electric charge, respectively.
Note that the lepton here is on-shell.

In fact, it is convenient to get the corrections from loop diagrams
to lepton MDMs and EDMs in terms of the effective Lagrangian method, if the masses of
internal lines are much heavier than the external lepton mass. Assuming external leptons
as well as photon are all off-shell, we expand the amplitude of the
corresponding triangle diagrams according to the external momenta of
leptons and photon. After matching between the effective theory and the full theory,
we can get all high dimension operators together
with their coefficients. As discussed in the section \ref{sec1}, it is enough to
retain only those dimension 6 operators in later calculations:
\begin{eqnarray}
&&{\cal O}_{_1}^\mp={1\over(4\pi)^2}\;\bar{l}\;(i/\!\!\!\!{\cal D})^3
\omega_\mp\;l\;,\nonumber\\
&&{\cal O}_{_2}^\mp={eQ_{_f}\over(4\pi)^2}\;\overline{(i{\cal D}_{_\mu}l)}
\gamma^\mu F\cdot\sigma\omega_\mp l\;,\nonumber\\
&&{\cal O}_{_3}^\mp={eQ_{_f}\over(4\pi)^2}\;\bar{l}F\cdot\sigma\gamma^\mu
\omega_\mp(i{\cal D}_{_\mu}l)\;,\nonumber\\
&&{\cal O}_{_4}^\mp={eQ_{_f}\over(4\pi)^2}\;\bar{l}(\partial^\mu F_{_{\mu\nu}})
\gamma^\nu\omega_\mp l\;,\nonumber\\
&&{\cal O}_{_5}^\mp={m_{_l}\over(4\pi)^2}\;\bar{l}\;(i/\!\!\!\!{\cal D})^2
\omega_\mp\;l\;,\nonumber\\
&&{\cal O}_{_6}^\mp={eQ_{_f}m_{_l}\over(4\pi)^2}\;\bar{l}\;F\cdot\sigma
\omega_\mp\;l\;,\nonumber\\
\label{ops}
\end{eqnarray}
with ${\cal D}_{_\mu}=\partial_{_\mu}+ieA_{_\mu}$ and $\omega_\mp=(1\mp\gamma_5)/2$.
When the equations of motion are applied to the incoming and outgoing leptons separately,
only the operators ${\cal O}_{_{2,3,6}}^\mp$ actually contribute to the MDMs and EDMs of leptons.
We will only present the Wilson coefficients of the operators ${\cal O}_{_{2,3,6}}^\mp$
in the effective Lagrangian in our following narration because of the reason mentioned above.

If the full theory is invariant under the combined transformation of charge
conjugation, parity and time reversal (CPT), the induced effective
theory preserves the symmetry after the heavy freedoms are integrated out.
The fact implies the Wilson coefficients of the operators ${\cal O}_{_{2,3,6}}^\mp$
satisfying the relations
\begin{eqnarray}
&&C_2^\mp=C_3^{\mp*},\;C_6^+=C_6^{-*}\;,
\label{cpt1}
\end{eqnarray}
where $C_i^\mp\;(i=1,2,\cdots,6)$ represent the Wilson coefficients
of the corresponding operators ${\cal O}_{_i}^\mp$ in the effective Lagrangian. After applying
the equations of motion to the external leptons, we find that the concerned terms
in the effective Lagrangian are transformed into
\begin{eqnarray}
&&C_2^\mp{\cal O}_{_2}^\mp+C_2^{\mp*}{\cal O}_{_3}^\mp+C_6^+{\cal O}_{_6}^+
+C_6^{+*}{\cal O}_{_6}^-
\nonumber\\
&&\hspace{-0.6cm}\Rightarrow
(C_2^++C_2^{-*}+C_6^+){\cal O}_{_6}^++(C_2^{+*}+C_2^-+C_6^{+*}){\cal O}_{_6}^-
\nonumber\\
&&\hspace{-0.6cm}=
{eQ_{_f}m_{_l}\over(4\pi)^2}\Big\{\Re(C_2^++C_2^{-*}+C_6^+)\;\bar{l}\;\sigma^{\mu\nu}\;l
+i\Im(C_2^++C_2^{-*}+C_6^+)\;\bar{l}\;\sigma^{\mu\nu}\gamma_5\;l\Big\}F_{\mu\nu}\;.
\label{cpt2}
\end{eqnarray}
Here, $\Re(\cdots)$ denotes the operation to take the real part of a complex number,
and $\Im(\cdots)$ denotes the operation to take the imaginary part of a complex number.
Applying Eq.(\ref{adm}) and Eq.(\ref{cpt2}), we finally get
\begin{eqnarray}
&&a_l={4Q_{_f}m_{_l}^2\over(4\pi)^2}\Re(C_2^++C_2^{-*}+C_6^+)\;,
\nonumber\\
&&d_l=-{2eQ_{_f}m_{_l}\over(4\pi)^2}\Im(C_2^++C_2^{-*}+C_6^+)\;.
\label{cpt3}
\end{eqnarray}
In other words, the MDM of lepton is proportional to real part of
the effective coupling $C_2^++C_2^{-*}+C_6^+$, as well as the EDM of lepton
is proportional to imaginary part of the effective coupling $C_2^++C_2^{-*}+C_6^+$.

Using the effective Lagrangian method,
we present the one-loop supersymmetric contribution to muon MDM
in \cite{Feng3} which coincides with the previous result in literature.
Since the complication of analysis at two-loop order, we will adopt below a
terminology where, for example, the "$\gamma h_k$" contribution means the sum of amplitude
from those triangle diagrams (indeed three triangles bound together), in which
a closed fermion (chargino/neutralino) loop is attached to  the virtual Higgs
and photon fields with a real photon attached in all possible ways to the
internal lines. Because the sum of amplitude from those "triangle" diagrams
corresponding to each "self-energy" obviously respects the Ward
identity requested by QED gauge symmetry, we can calculate the
contributions of all the "self-energies" separately. Taking the same steps which
we did in our earlier works \cite{Feng1,Feng2,Feng3}, we obtain the
effective Lagrangian that originates from the self energy diagrams in Fig.\ref{fig1}.
In the bare effective Lagrangian from the 'WW' and 'ZZ' contributions,
the ultraviolet divergence caused by divergent sub-diagrams can be subtracted safely
in on-mass-shell scheme \cite{onshell}. Now, we present the effective Lagrangian
corresponding to the diagrams in Fig.\ref{fig1} respectively.

\subsection{The effective Lagrangian from $\gamma h_k\;\;(k=1,2,3)$ and
$\gamma G_0$ sector}
\indent\indent
As a closed chargino loop is attached to the virtual neutral Higgs and photon
fields, a real photon can be emitted from either the virtual lepton or the virtual charginos
in the self energy diagram. When a real photon is emitted
from the virtual charginos, the corresponding "triangle" diagrams
belong to the typical two-loop Bar-Zee-type diagrams \cite{Barr-Zee}.
Within the framework of CP violating MSSM,
the contributions from two-loop Bar-Zee-type diagrams to the EDMs of those light fermions
are discussed extensively in literature \cite{BZ-mssm}.
When a real photon is attached to the internal standard fermion,
the correction from corresponding triangle diagram to
the effective Lagrangian is zero because of the Furry theorem,
this point is also verified through a strict analysis.
The corresponding effective Lagrangian from this sector is written as
\begin{eqnarray}
&&{\cal L}_{_{\gamma h_k}}=
{e^4(Z_{_H})_{_{1k}}\over2\sqrt{2}(4\pi)^2s_{_{\rm w}}^2\Lambda^2\cos\beta}
\Bigg\{\Re({\cal H}_{ii}^k)\Big({x_{_{\chi_i^\pm}}\over x_{_{\rm w}}}\Big)^{1/2}
T_1(x_{_{h_k}},x_{_{\chi_i^\pm}},x_{_{\chi_i^\pm}})
\Big({\cal O}_{_6}^++{\cal O}_{_6}^-\Big)
\nonumber\\
&&\hspace{1.2cm}
+i\Im({\cal H}_{ii}^k)\Big({x_{_{\chi_i^\pm}}\over x_{_{\rm w}}}\Big)^{1/2}
T_2(x_{_{h_k}},x_{_{\chi_i^\pm}},x_{_{\chi_i^\pm}})
\Big({\cal O}_{_6}^+-{\cal O}_{_6}^-\Big)\Bigg\}
\nonumber\\
&&\hspace{1.2cm}
-{e^4(Z_{_H})_{_{3k}}\tan\beta\over2\sqrt{2}(4\pi)^2s_{_{\rm w}}^2\Lambda^2}
\Bigg\{\Re({\cal A}_{ii}^k)
\Big({x_{_{\chi_i^\pm}}\over x_{_{\rm w}}}\Big)^{1/2}
T_2(x_{_{h_k}},x_{_{\chi_i^\pm}},x_{_{\chi_i^\pm}})
\Bigg]\Big({\cal O}_{_6}^++{\cal O}_{_6}^-\Big)
\nonumber\\
&&\hspace{1.2cm}
-i\Im({\cal A}_{ii}^k)\Big({x_{_{\chi_i^\pm}}\over x_{_{\rm w}}}\Big)^{1/2}
T_1(x_{_{h_k}},x_{_{\chi_i^\pm}},x_{_{\chi_i^\pm}})
\Big({\cal O}_{_6}^+-{\cal O}_{_6}^-\Big)\Bigg\}\;.
\label{gamma-h}
\end{eqnarray}
with
\begin{eqnarray}
&&{\cal H}_{ij}^k=(U_{_R}^\dagger)_{_{i1}}(U_{_L})_{_{2j}}(Z_{_H})_{_{1k}}
+(U_{_R}^\dagger)_{_{i2}}(U_{_L})_{_{1j}}(Z_{_H})_{_{2k}}\;,
\nonumber\\
&&{\cal A}_{ij}^k=\Big((U_{_R}^\dagger)_{_{i1}}(U_{_L})_{_{2j}}\sin\beta
+(U_{_R}^\dagger)_{_{i2}}(U_{_L})_{_{1j}}\cos\beta\Big)(Z_{_H})_{_{3k}}
\;,\;(i,j=1,2)\;.
\label{coupling-h}
\end{eqnarray}
Where the two unitary matrices $U_{_{L,R}}$ denote the left- and right-mixing
matrices of charginos, $\Lambda$ denotes the energy scale of new physics,
and $x_i=m_i^2/\Lambda^2$ respectively. We adopt the abbreviations:
$c_{_{\rm w}}=\cos\theta_{_{\rm w}},\;s_{_{\rm w}}
=\sin\theta_{_{\rm w}},\;$ where $\theta_{_{\rm w}}$ is the Weinberg angle.
The concrete expressions of $T_{1,2}$ can be found in appendix.

Accordingly, the lepton MDMs and EDMs from $\gamma h_k$ sector are written as
\begin{eqnarray}
&&a_l^{\gamma h_k}={\sqrt{2}e^4Q_{_f}m_{_l}^2(Z_{_H})_{_{1k}}\over(4\pi)^4
s_{_{\rm w}}^2\Lambda^2\cos\beta}\Re({\cal H}_{ii}^k)\Big({x_{_{\chi_i^\pm}}
\over x_{_{\rm w}}}\Big)^{1/2}T_1(x_{_{h_k}},x_{_{\chi_i^\pm}},x_{_{\chi_i^\pm}})
\nonumber\\
&&\hspace{1.2cm}
-{\sqrt{2}e^4Q_{_f}m_{_l}^2(Z_{_H})_{_{3k}}\tan\beta\over(4\pi)^4
s_{_{\rm w}}^2\Lambda^2}\Re({\cal A}_{ii}^k)\Big({x_{_{\chi_i^\pm}}
\over x_{_{\rm w}}}\Big)^{1/2}T_2(x_{_{h_k}},x_{_{\chi_i^\pm}},x_{_{\chi_i^\pm}})
\;,\nonumber\\
&&d_l^{\gamma h_k}=-{e^5Q_{_f}m_{_l}(Z_{_H})_{_{1k}}\over\sqrt{2}
(4\pi)^4s_{_{\rm w}}^2\Lambda^2\cos\beta}
\Im({\cal H}_{ii}^k)\Big({x_{_{\chi_i^\pm}}\over x_{_{\rm w}}}\Big)^{1/2}
T_2(x_{_{h_k}},x_{_{\chi_i^\pm}},x_{_{\chi_i^\pm}})
\nonumber\\
&&\hspace{1.2cm}
-{e^5Q_{_f}m_{_l}(Z_{_H})_{_{3k}}\tan\beta\over\sqrt{2}
(4\pi)^4s_{_{\rm w}}^2\Lambda^2}
\Im({\cal A}_{ii}^k)\Big({x_{_{\chi_i^\pm}}\over x_{_{\rm w}}}\Big)^{1/2}
T_1(x_{_{h_k}},x_{_{\chi_i^\pm}},x_{_{\chi_i^\pm}})\;,
\label{MD-gamma-h}
\end{eqnarray}
which are enhanced by large $\tan\beta$. Note here that the corrections
from this sector to the MDM of lepton
depend on a linear combination of real parts of the effective couplings
${\cal H}_{ii}^k$ and  ${\cal A}_{ii}^k$, and the corrections from this sector
to the EDM of lepton depend on a linear combination of imaginary parts
of the effective couplings ${\cal H}_{ii}^k$ and ${\cal A}_{ii}^k$.
In the limit $x_{_{\chi_i^\pm}}\gg x_{_{h_k}}$, the above expressions can
be simplified as
\begin{eqnarray}
&&a_l^{\gamma h_k}=-{\sqrt{2}e^4Q_{_f}m_{_l}^2(Z_{_H})_{_{1k}}\over(4\pi)^4
s_{_{\rm w}}^2\Lambda^2\cos\beta}\Re({\cal H}_{ii}^k)\Big({x_{_{\chi_i^\pm}}
\over x_{_{\rm w}}}\Big)^{1/2}\lim\limits_{x_{_{\chi_j^\pm}}\rightarrow x_{_{\chi_i^\pm}}}
{\partial\over\partial x_{_{\chi_j^\pm}}}\varphi_1(x_{_{\chi_i^\pm}},
x_{_{\chi_j^\pm}})
\nonumber\\
&&\hspace{1.2cm}
-{\sqrt{2}e^4Q_{_f}m_{_l}^2(Z_{_H})_{_{3k}}\tan\beta\over(4\pi)^4
s_{_{\rm w}}^2\Lambda^2}\Re({\cal A}_{ii}^k)\Big({x_{_{\chi_i^\pm}}
\over x_{_{\rm w}}}\Big)^{1/2}\Big[{\ln x_{_{h_k}}\over x_{_{\chi_i^\pm}}}
+\lim\limits_{x_{_{\chi_j^\pm}}\rightarrow x_{_{\chi_i^\pm}}}
{\partial\over\partial x_{_{\chi_j^\pm}}}\varphi_1(x_{_{\chi_i^\pm}},
x_{_{\chi_j^\pm}})\Big]
\;,\nonumber\\
&&d_l^{\gamma h_k}=-{e^5Q_{_f}m_{_l}(Z_{_H})_{_{1k}}\over\sqrt{2}
(4\pi)^4s_{_{\rm w}}^2\Lambda^2\cos\beta}
\Im({\cal H}_{ii}^k)\Big({x_{_{\chi_i^\pm}}\over x_{_{\rm w}}}\Big)^{1/2}
\Big[{\ln x_{_{h_k}}\over x_{_{\chi_i^\pm}}}
+\lim\limits_{x_{_{\chi_j^\pm}}\rightarrow x_{_{\chi_i^\pm}}}
{\partial\over\partial x_{_{\chi_j^\pm}}}\varphi_1(x_{_{\chi_i^\pm}},
x_{_{\chi_j^\pm}})\Big]
\nonumber\\
&&\hspace{1.2cm}
+{e^5Q_{_f}m_{_l}(Z_{_H})_{_{3k}}\tan\beta\over\sqrt{2}
(4\pi)^4s_{_{\rm w}}^2\Lambda^2}
\Im({\cal A}_{ii}^k)\Big({x_{_{\chi_i^\pm}}\over x_{_{\rm w}}}\Big)^{1/2}
\lim\limits_{x_{_{\chi_j^\pm}}\rightarrow x_{_{\chi_i^\pm}}}
{\partial\over\partial x_{_{\chi_j^\pm}}}\varphi_1(x_{_{\chi_i^\pm}},
x_{_{\chi_j^\pm}})\;.
\label{MD-gamma-h1}
\end{eqnarray}

Similarly, we can formulate the corrections from $\gamma G_0$ sector to
the effective Lagrangian as
\begin{eqnarray}
&&{\cal L}_{_{\gamma G}}=
{e^4\over2\sqrt{2}(4\pi)^2s_{_{\rm w}}^2\Lambda^2}\Bigg\{\Re({\cal B}_{ii})
\Big({x_{_{\chi_i^\pm}}\over x_{_{\rm w}}}\Big)^{1/2}
T_2(x_{_{\rm z}},x_{_{\chi_i^\pm}},x_{_{\chi_i^\pm}})
\Bigg]\Big({\cal O}_{_6}^++{\cal O}_{_6}^-\Big)
\nonumber\\
&&\hspace{1.2cm}
-i\Im({\cal B}_{ii})\Big({x_{_{\chi_i^\pm}}\over x_{_{\rm w}}}\Big)^{1/2}
T_1(x_{_{\rm z}},x_{_{\chi_i^\pm}},x_{_{\chi_i^\pm}})
\Big({\cal O}_{_6}^+-{\cal O}_{_6}^-\Big)\Bigg\}\;,
\label{gamma-G}
\end{eqnarray}
with
\begin{eqnarray}
&&{\cal B}_{ij}=-(U_{_R}^\dagger)_{_{i1}}(U_{_L})_{_{2j}}\cos\beta
+(U_{_R}^\dagger)_{_{i2}}(U_{_L})_{_{1j}}\sin\beta\;,\;(i,j=1,2)\;.
\label{coupling-neutrnal-Goldstone}
\end{eqnarray}

Correspondingly, the corrections to the lepton MDMs and EDMs from this
sector are:
\begin{eqnarray}
&&a_l^{\gamma G}={\sqrt{2}e^4Q_{_f}m_{_l}^2\over(4\pi)^4s_{_{\rm w}}^2\Lambda^2}
\Re({\cal B}_{ii})\Big({x_{_{\chi_i^\pm}}\over x_{_{\rm w}}}\Big)^{1/2}
T_2(x_{_{\rm z}},x_{_{\chi_i^\pm}},x_{_{\chi_i^\pm}})\;,
\nonumber\\
&&d_l^{\gamma G}={e^5Q_{_f}m_{_l}\over\sqrt{2}(4\pi)^4s_{_{\rm w}}^2\Lambda^2}
\Im({\cal B}_{ii})\Big({x_{_{\chi_i^\pm}}\over x_{_{\rm w}}}\Big)^{1/2}
T_1(x_{_{\rm z}},x_{_{\chi_i^\pm}},x_{_{\chi_i^\pm}})\;.
\label{MD-gamma-G}
\end{eqnarray}
The corrections from this sector to the MDM of lepton are proportional
to real parts of the effective couplings ${\cal B}_{ii}$, and the corrections
from this sector to  the EDM of lepton are proportional to imaginary parts
of the effective couplings ${\cal B}_{ii}$, separately.
In the limit $x_{_{\chi_i^\pm}}\gg x_{_{\rm z}}$, we have
\begin{eqnarray}
&&a_l^{\gamma G}={\sqrt{2}e^4Q_{_f}m_{_l}^2\over(4\pi)^4s_{_{\rm w}}^2\Lambda^2}
\Re({\cal B}_{ii})\Big({x_{_{\chi_i^\pm}}\over x_{_{\rm w}}}\Big)^{1/2}
\Big[{\ln x_{_{\rm z}}\over x_{_{\chi_i^\pm}}}
+\lim\limits_{x_{_{\chi_j^\pm}}\rightarrow x_{_{\chi_i^\pm}}}
{\partial\over\partial x_{_{\chi_j^\pm}}}\varphi_1(x_{_{\chi_i^\pm}},
x_{_{\chi_j^\pm}})\Big]\;,
\nonumber\\
&&d_l^{\gamma G}={e^5Q_{_f}m_{_l}\over\sqrt{2}(4\pi)^4s_{_{\rm w}}^2\Lambda^2}
\Im({\cal B}_{ii})\Big({x_{_{\chi_i^\pm}}\over x_{_{\rm w}}}\Big)^{1/2}
\lim\limits_{x_{_{\chi_j^\pm}}\rightarrow x_{_{\chi_i^\pm}}}
{\partial\over\partial x_{_{\chi_j^\pm}}}\varphi_1(x_{_{\chi_i^\pm}},
x_{_{\chi_j^\pm}})\;.
\label{MD-gamma-G1}
\end{eqnarray}

Using the concrete expression of $\varphi_1(x,y)$ presented in appendix,
one can verify easily that the corrections to the lepton MDMs and EDMs
from the sectors are suppressed by the masses of charginos as
$m_{_{\chi_i^\pm}}\gg m_{_{h_k}},\;m_{_{\rm z}}\;(i=1,\;2)$.

\subsection{The effective Lagrangian from $Zh_k$ ($ZG_0$) sector}
\indent\indent
As a closed chargino loop is attached to the virtual Higgs and $Z$ gauge boson
fields, a real photon can be attached to either the virtual lepton or the virtual charginos
in the self energy diagram. When a real photon is attached to the virtual lepton,
the corresponding amplitude only modifies the Wilson coefficients of the
operators ${\cal O}_{_5}^\pm$ in the effective Lagrangian after the heavy freedoms
are integrated out. In other words, this triangle diagram does not contribute to the
lepton MDMs and EDMs. A real photon can be only attached to the virtual lepton
as the closed loop is composed of neutralinos,
the corresponding triangle diagram does not affect the theoretical predictions
on the lepton MDMs and EDMs for the same reason.
Considering the points above, we formulate the contributions from $Zh_0$ sector
to the effective Lagrangian as
\begin{eqnarray}
&&{\cal L}_{_{Zh_k}}=
-{e^4(Z_{_H})_{_{1k}}\over16\sqrt{2}(4\pi)^2s_{_{\rm w}}^4c_{_{\rm w}}^2Q_{_f}\Lambda^2\cos\beta}
(T_{_f}^Z-2Q_{_f}s_{_{\rm w}}^2)\Bigg\{\Big({x_{_{\chi_j^\pm}}\over x_{_{\rm w}}}\Big)^{1/2}
\Big[(4+2\ln x_{_{\chi_j^\pm}})
\nonumber\\
&&\hspace{1.2cm}\times
\varrho_{_{0,1}}(x_{_{\rm z}},x_{_{h_k}})
+F_1(x_{_{\rm z}},x_{_{h_k}},x_{_{\chi_i^\pm}},x_{_{\chi_j^\pm}})\Big]
\Re\Big({\cal H}_{_{ji}}^k\xi^L_{_{ij}}+{\cal H}^{k,\dagger}_{_{ji}}
\xi^R_{_{ij}}\Big)({\cal O}_{_6}^++{\cal O}_{_6}^-)
\nonumber\\
&&\hspace{1.2cm}
+i\Big({x_{_{\chi_j^\pm}}\over x_{_{\rm w}}}\Big)^{1/2}
\Big[-2(\ln x_{_{\chi_i^\pm}}-\ln x_{_{\chi_j^\pm}})
\varrho_{_{0,1}}(x_{_{\rm z}},x_{_{h_k}})
+F_1(x_{_{\rm z}},x_{_{h_k}},x_{_{\chi_i^\pm}},x_{_{\chi_j^\pm}})
\nonumber\\
&&\hspace{1.2cm}
+F_2(x_{_{\rm z}},x_{_{h_k}},x_{_{\chi_j^\pm}},x_{_{\chi_i^\pm}})\Big]
\Im\Big({\cal H}_{_{ji}}^k\xi^L_{_{ij}}-{\cal H}^{k,\dagger}_{_{ji}}
\xi^R_{_{ij}}\Big)({\cal O}_{_6}^--{\cal O}_{_6}^+)\Bigg\}
\nonumber\\
&&\hspace{1.2cm}
+{e^4(Z_{_H})_{_{3k}}\tan\beta\over16\sqrt{2}(4\pi)^2s_{_{\rm w}}^4c_{_{\rm w}}^2
Q_{_f}\Lambda^2}(T_{_f}^Z-2Q_{_f}s_{_{\rm w}}^2)
\Bigg\{-i\Big({x_{_{\chi_j^\pm}}\over x_{_{\rm w}}}\Big)^{1/2}
\Big[2(2+\ln x_{_{\chi_j^\pm}})\varrho_{_{0,1}}(x_{_{\rm z}},x_{_{h_k}})
\nonumber\\
&&\hspace{1.2cm}
+F_1(x_{_{\rm z}},x_{_{h_k}},x_{_{\chi_i^\pm}},x_{_{\chi_j^\pm}})\Big]
\Im\Big({\cal A}_{_{ji}}^k\xi^L_{_{ij}}+{\cal A}^{k,\dagger}_{_{ji}}
\xi^R_{_{ij}}\Big)({\cal O}_{_6}^--{\cal O}_{_6}^+)
\nonumber\\
&&\hspace{1.2cm}
+\Big({x_{_{\chi_j^\pm}}\over x_{_{\rm w}}}\Big)^{1/2}
\Big[-2(\ln x_{_{\chi_i^\pm}}-\ln x_{_{\chi_j^\pm}})
\varrho_{_{0,1}}(x_{_{\rm z}},x_{_{h_k}})
+F_1(x_{_{\rm z}},x_{_{h_k}},x_{_{\chi_i^\pm}},x_{_{\chi_j^\pm}})
\nonumber\\
&&\hspace{1.2cm}
+F_2(x_{_{\rm z}},x_{_{h_k}},x_{_{\chi_j^\pm}},x_{_{\chi_i^\pm}})\Big]
\Re\Big({\cal A}_{_{ji}}^k\xi^L_{_{ij}}-{\cal A}^{k,\dagger}_{_{ji}}
\xi^R_{_{ij}}\Big)({\cal O}_{_6}^-+{\cal O}_{_6}^+)\Bigg\}
+\cdots
\label{zh}
\end{eqnarray}
with
\begin{eqnarray}
&&\xi^L_{ij}=2\delta_{ij}\cos2\theta_{_{\rm w}}
+(U_{_L}^\dagger)_{_{i1}}(U_{_L})_{_{1j}}
\;,\nonumber\\
&&\xi^R_{ij}=2\delta_{ij}\cos2\theta_{_{\rm w}}
+(U_{_R}^\dagger)_{_{i1}}(U_{_R})_{_{1j}}\;,\;(i,j=1,2)\;,
\label{coupling-xi}
\end{eqnarray}
where the concrete expressions of the functions $\varrho_{_{i,j}}(x_1,x_2),\;
F_{1,2}(x_1,x_2,x_3,x_4)$ are listed in appendix. Additional, $T_{_f}^Z$
is the isospin of lepton, and $Q_{_f}$ is the electric charge of lepton, respectively.
Using Eq.\ref{zh}, we get the corrections to the lepton MDMs and EDMs from $Zh_k$ sector as
\begin{eqnarray}
&&a_l^{Zh_k}=-{e^4m_{_l}^2(Z_{_H})_{_{1k}}\over4\sqrt{2}(4\pi)^4s_{_{\rm w}}^4
c_{_{\rm w}}^2\Lambda^2\cos\beta}
(T_{_f}^Z-2Q_{_f}s_{_{\rm w}}^2)\Big({x_{_{\chi_j^\pm}}\over x_{_{\rm w}}}\Big)^{1/2}
\Big[2(2+\ln x_{_{\chi_j^\pm}})\varrho_{_{i,j}}(x_{_{\rm z}},x_{_{h_k}})
\nonumber\\
&&\hspace{1.2cm}
+F_1(x_{_{\rm z}},x_{_{h_k}},x_{_{\chi_i^\pm}},x_{_{\chi_j^\pm}})\Big]
\Re\Big({\cal H}_{_{ji}}^k\xi^L_{_{ij}}+{\cal H}^{k,\dagger}_{_{ji}}
\xi^R_{_{ij}}\Big)
\nonumber\\
&&\hspace{1.2cm}
+{e^4m_{_l}^2(Z_{_H})_{_{3k}}\tan\beta\over4\sqrt{2}(4\pi)^4s_{_{\rm w}}^4
c_{_{\rm w}}^2\Lambda^2}
(T_{_f}^Z-2Q_{_f}s_{_{\rm w}}^2)\Big({x_{_{\chi_j^\pm}}\over x_{_{\rm w}}}\Big)^{1/2}
\Big[-2(\ln x_{_{\chi_i^\pm}}-\ln x_{_{\chi_j^\pm}})
\varrho_{_{0,1}}(x_{_{\rm z}},x_{_{h_k}})
\nonumber\\
&&\hspace{1.2cm}
+F_1(x_{_{\rm z}},x_{_{h_k}},x_{_{\chi_i^\pm}},x_{_{\chi_j^\pm}})
+F_2(x_{_{\rm z}},x_{_{h_k}},x_{_{\chi_j^\pm}},x_{_{\chi_i^\pm}})\Big]
\Re\Big({\cal A}_{_{ji}}^k\xi^L_{_{ij}}-{\cal A}^{k,\dagger}_{_{ji}}
\xi^R_{_{ij}}\Big)\;,
\nonumber\\
&&d_l^{Zh_k}={e^5m_{_l}(Z_{_H})_{_{1k}}\over8\sqrt{2}(4\pi)^4s_{_{\rm w}}^4
c_{_{\rm w}}^2\Lambda^2\cos\beta}
(T_{_f}^Z-2Q_{_f}s_{_{\rm w}}^2)\Big({x_{_{\chi_j^\pm}}\over x_{_{\rm w}}}\Big)^{1/2}
\Big[2(\ln x_{_{\chi_i^\pm}}-\ln x_{_{\chi_j^\pm}})
\varrho_{_{0,1}}(x_{_{\rm z}},x_{_{h_k}})
\nonumber\\
&&\hspace{1.2cm}
-F_1(x_{_{\rm z}},x_{_{h_k}},x_{_{\chi_i^\pm}},x_{_{\chi_j^\pm}})
-F_2(x_{_{\rm z}},x_{_{h_k}},x_{_{\chi_j^\pm}},x_{_{\chi_i^\pm}})\Big]
\Im\Big({\cal H}_{_{ji}}^k\xi^L_{_{ij}}-{\cal H}^{k,\dagger}_{_{ji}}
\xi^R_{_{ij}}\Big)
\nonumber\\
&&\hspace{1.2cm}
-{e^5m_{_l}(Z_{_H})_{_{3k}}\tan\beta\over8\sqrt{2}(4\pi)^4s_{_{\rm w}}^4
c_{_{\rm w}}^2\Lambda^2}(T_{_f}^Z-2Q_{_f}s_{_{\rm w}}^2)\Big({x_{_{\chi_j^\pm}}\over
x_{_{\rm w}}}\Big)^{1/2}
\Big[2(2+\ln x_{_{\chi_j^\pm}})\varrho_{_{i,j}}(x_{_{\rm z}},x_{_{h_k}})
\nonumber\\
&&\hspace{1.2cm}
+F_1(x_{_{\rm z}},x_{_{h_k}},x_{_{\chi_i^\pm}},x_{_{\chi_j^\pm}})\Big]
\Im\Big({\cal A}_{_{ji}}^k\xi^L_{_{ij}}+{\cal A}^{k,\dagger}_{_{ji}}
\xi^R_{_{ij}}\Big)\;.
\label{MD-z-h}
\end{eqnarray}
The above equations contain the suppression factor $1-4s_{_{\rm w}}^2$ because $Q_{_f}=-1$
and $T_{_f}^Z=-1/2$ for charged leptons. The corrections from this sector to the MDM of lepton
are decided by a linear combination of real parts of the effective couplings
${\cal H}_{_{ji}}^k\xi^L_{_{ij}}+{\cal H}^{k,\dagger}_{_{ji}}\xi^R_{_{ij}}$ and
${\cal A}_{_{ji}}^k\xi^L_{_{ij}}-{\cal A}^{k,\dagger}_{_{ji}}\xi^R_{_{ij}}$,
and the corrections from this sector to the EDM of lepton
are decided by a linear combination of imaginary parts of the effective couplings
${\cal H}_{_{ji}}^k\xi^L_{_{ij}}-{\cal H}^{k,\dagger}_{_{ji}}\xi^R_{_{ij}}$ and
${\cal A}_{_{ji}}^k\xi^L_{_{ij}}+{\cal A}^{k,\dagger}_{_{ji}}\xi^R_{_{ij}}$.
In the limit $x_{_{\chi_i^\pm}},\;x_{_{\chi_j^\pm}}\gg x_{_{\rm z}},\;x_{_{h_k}}$,
Eq.\ref{MD-z-h} can be approximated as
\begin{eqnarray}
&&a_l^{Zh_k}=-{e^4m_{_l}^2(Z_{_H})_{_{1k}}\over4\sqrt{2}(4\pi)^4s_{_{\rm w}}^4
c_{_{\rm w}}^2\Lambda^2\cos\beta}
(T_{_f}^Z-2Q_{_f}s_{_{\rm w}}^2)\Big({x_{_{\chi_j^\pm}}\over x_{_{\rm w}}}\Big)^{1/2}
\Big[{\partial\varphi_1\over\partial x_{_{\chi_j^\pm}}}(x_{_{\chi_i^\pm}}
,x_{_{\chi_j^\pm}})
\nonumber\\
&&\hspace{1.2cm}
-{2-2x_{_{\chi_i^\pm}}\varrho_{_{0,1}}(x_{_{\chi_i^\pm}},x_{_{\chi_j^\pm}})
\over x_{_{\chi_i^\pm}}-x_{_{\chi_j^\pm}}}\cdot
\varrho_{_{1,1}}(x_{_{\rm z}},x_{_{h_k}})\Big]
\Re\Big({\cal H}_{_{ji}}^k\xi^L_{_{ij}}+{\cal H}^{k,\dagger}_{_{ji}}
\xi^R_{_{ij}}\Big)
\nonumber\\
&&\hspace{1.2cm}
+{e^4m_{_l}^2(Z_{_H})_{_{3k}}\tan\beta\over4\sqrt{2}(4\pi)^4s_{_{\rm w}}^4
c_{_{\rm w}}^2\Lambda^2}
(T_{_f}^Z-2Q_{_f}s_{_{\rm w}}^2)\Big({x_{_{\chi_j^\pm}}\over x_{_{\rm w}}}\Big)^{1/2}
\Big[\Big({\partial\varphi_1\over\partial x_{_{\chi_i^\pm}}}
+{\partial\varphi_1\over\partial x_{_{\chi_j^\pm}}}\Big)
(x_{_{\chi_i^\pm}},x_{_{\chi_j^\pm}})
\nonumber\\
&&\hspace{1.2cm}
+2\varrho_{_{0,1}}(x_{_{\chi_i^\pm}},x_{_{\chi_j^\pm}})
\varrho_{_{1,1}}(x_{_{\rm z}},x_{_{h_k}})\Big]
\Re\Big({\cal A}_{_{ji}}^k\xi^L_{_{ij}}-{\cal A}^{k,\dagger}_{_{ji}}
\xi^R_{_{ij}}\Big)\;,
\nonumber\\
&&d_l^{Zh_k}=-{e^5m_{_l}(Z_{_H})_{_{1k}}\over8\sqrt{2}(4\pi)^4s_{_{\rm w}}^4
c_{_{\rm w}}^2\Lambda^2\cos\beta}
(T_{_f}^Z-2Q_{_f}s_{_{\rm w}}^2)\Big({x_{_{\chi_j^\pm}}\over x_{_{\rm w}}}\Big)^{1/2}
\Big[\Big({\partial\varphi_1\over\partial x_{_{\chi_i^\pm}}}
+{\partial\varphi_1\over\partial x_{_{\chi_j^\pm}}}\Big)
(x_{_{\chi_i^\pm}},x_{_{\chi_j^\pm}})
\nonumber\\
&&\hspace{1.2cm}
+2\varrho_{_{0,1}}(x_{_{\chi_i^\pm}},x_{_{\chi_j^\pm}})
\varrho_{_{1,1}}(x_{_{\rm z}},x_{_{h_k}})\Big]
\Im\Big({\cal H}_{_{ji}}^k\xi^L_{_{ij}}-{\cal H}^{k,\dagger}_{_{ji}}
\xi^R_{_{ij}}\Big)
\nonumber\\
&&\hspace{1.2cm}
-{e^5m_{_l}(Z_{_H})_{_{3k}}\tan\beta\over8\sqrt{2}(4\pi)^4s_{_{\rm w}}^4
c_{_{\rm w}}^2\Lambda^2}(T_{_f}^Z-2Q_{_f}s_{_{\rm w}}^2)\Big({x_{_{\chi_j^\pm}}\over
x_{_{\rm w}}}\Big)^{1/2}
\Big[{\partial\varphi_1\over\partial x_{_{\chi_j^\pm}}}(x_{_{\chi_i^\pm}}
,x_{_{\chi_j^\pm}})
\nonumber\\
&&\hspace{1.2cm}
-{2-2x_{_{\chi_i^\pm}}\varrho_{_{0,1}}(x_{_{\chi_i^\pm}},x_{_{\chi_j^\pm}})
\over x_{_{\chi_i^\pm}}-x_{_{\chi_j^\pm}}}\cdot
\varrho_{_{1,1}}(x_{_{\rm z}},x_{_{h_k}})\Big]
\Im\Big({\cal A}_{_{ji}}^k\xi^L_{_{ij}}+{\cal A}^{k,\dagger}_{_{ji}}
\xi^R_{_{ij}}\Big)\;.
\label{MD-z-h1}
\end{eqnarray}

Similarly, the contribution from $ZG_0$ sector to the effective Lagrangian is
\begin{eqnarray}
&&{\cal L}_{_{ZG_0}}=
-{e^4\over16\sqrt{2}(4\pi)^2s_{_{\rm w}}^4c_{_{\rm w}}^2Q_{_f}\Lambda^2}
\Bigg\{-i\Big({x_{_{\chi_j^\pm}}\over x_{_{\rm w}}}\Big)^{1/2}
\Big[{2\over x_{_{\rm z}}}(2+\ln x_{_{\chi_j^\pm}})
+F_1(x_{_{\rm z}}, x_{_{\rm z}},x_{_{\chi_i^\pm}},x_{_{\chi_j^\pm}})\Big]
\nonumber\\
&&\hspace{1.2cm}\times
\Im\Big({\cal B}_{_{ji}}\xi^L_{_{ij}}+{\cal B}^\dagger_{_{ji}}
\xi^R_{_{ij}}\Big)
(T_{_f}^Z-2Q_{_f}s_{_{\rm w}}^2)({\cal O}_{_6}^--{\cal O}_{_6}^+)
\nonumber\\
&&\hspace{1.2cm}
+\Big({x_{_{\chi_j^\pm}}\over x_{_{\rm w}}}\Big)^{1/2}
\Big[-{2\over  x_{_{\rm z}}}(\ln x_{_{\chi_i^\pm}}-\ln x_{_{\chi_j^\pm}})
+F_1(x_{_{\rm z}},x_{_{\rm z}},x_{_{\chi_i^\pm}},x_{_{\chi_j^\pm}})
+F_2(x_{_{\rm z}},x_{_{\rm z}},x_{_{\chi_j^\pm}},x_{_{\chi_i^\pm}})\Big]
\nonumber\\
&&\hspace{1.2cm}\times
\Re\Big({\cal B}_{_{ji}}\xi^L_{_{ij}}-{\cal B}^\dagger_{_{ji}}
\xi^R_{_{ij}}\Big)(T_{_f}^Z-2Q_{_f}s_{_{\rm w}}^2)({\cal O}_{_6}^-+{\cal O}_{_6}^+)\Bigg\}
+\cdots\;,
\label{zG}
\end{eqnarray}
and the contributions to the lepton MDMs and EDMs are:
\begin{eqnarray}
&&a_l^{ZG}=-{e^4m_{_l}^2\over4\sqrt{2}(4\pi)^4s_{_{\rm w}}^4c_{_{\rm w}}^2\Lambda^2}
(T_{_f}^Z-2Q_{_f}s_{_{\rm w}}^2)\Big({x_{_{\chi_j^\pm}}\over x_{_{\rm w}}}\Big)^{1/2}
\Big[-{2\over  x_{_{\rm z}}}(\ln x_{_{\chi_i^\pm}}-\ln x_{_{\chi_j^\pm}})
\nonumber\\
&&\hspace{1.2cm}
+F_1(x_{_{\rm z}},x_{_{\rm z}},x_{_{\chi_i^\pm}},x_{_{\chi_j^\pm}})
+F_2(x_{_{\rm z}},x_{_{\rm z}},x_{_{\chi_j^\pm}},x_{_{\chi_i^\pm}})\Big]
\Re\Big({\cal B}_{_{ji}}\xi^L_{_{ij}}-{\cal B}^\dagger_{_{ji}}
\xi^R_{_{ij}}\Big)\;,
\nonumber\\
&&d_l^{ZG}={e^5m_{_l}\over8\sqrt{2}(4\pi)^4s_{_{\rm w}}^4c_{_{\rm w}}^2\Lambda^2}
(T_{_f}^Z-2Q_{_f}s_{_{\rm w}}^2)\Big({x_{_{\chi_j^\pm}}\over x_{_{\rm w}}}\Big)^{1/2}
\Big[{2\over x_{_{\rm z}}}(2+\ln x_{_{\chi_j^\pm}})
\nonumber\\
&&\hspace{1.2cm}
+F_1(x_{_{\rm z}}, x_{_{\rm z}},x_{_{\chi_i^\pm}},x_{_{\chi_j^\pm}})\Big]
\Im\Big({\cal B}_{_{ji}}\xi^L_{_{ij}}+{\cal B}^\dagger_{_{ji}}
\xi^R_{_{ij}}\Big)\;.
\label{MD-z-G}
\end{eqnarray}
Here, the corrections from this sector to the MDM of lepton are proportional
to real parts of the effective couplings ${\cal B}_{_{ji}}\xi^L_{_{ij}}
-{\cal B}^\dagger_{_{ji}}\xi^R_{_{ij}}$, and the corrections from this sector
to the EDM of lepton are proportional to imaginary parts of the effective
couplings ${\cal B}_{_{ji}}\xi^L_{_{ij}}+{\cal B}^\dagger_{_{ji}}\xi^R_{_{ij}}$.
When $x_{_{\chi_i^\pm}},\;x_{_{\chi_j^\pm}}\gg x_{_{\rm z}}$,
Eq.\ref{MD-z-G} can be approached by
\begin{eqnarray}
&&a_l^{ZG}=-{e^4m_{_l}^2\over4\sqrt{2}(4\pi)^4s_{_{\rm w}}^4c_{_{\rm w}}^2\Lambda^2}
(T_{_f}^Z-2Q_{_f}s_{_{\rm w}}^2)\Big({x_{_{\chi_j^\pm}}\over x_{_{\rm w}}}\Big)^{1/2}
\Big[\Big({\partial\varphi_1\over\partial x_{_{\chi_i^\pm}}}
+{\partial\varphi_1\over\partial x_{_{\chi_j^\pm}}}\Big)
(x_{_{\chi_i^\pm}},x_{_{\chi_j^\pm}})
\nonumber\\
&&\hspace{1.2cm}
+2(1+\ln x_{_{\rm z}})\varrho_{_{0,1}}(x_{_{\chi_i^\pm}},x_{_{\chi_j^\pm}})\Big]
\Re\Big({\cal B}_{_{ji}}\xi^L_{_{ij}}-{\cal B}^\dagger_{_{ji}}
\xi^R_{_{ij}}\Big)\;,
\nonumber\\
&&d_l^{ZG}={e^5m_{_l}\over8\sqrt{2}(4\pi)^4s_{_{\rm w}}^4c_{_{\rm w}}^2\Lambda^2}
(T_{_f}^Z-2Q_{_f}s_{_{\rm w}}^2)\Big({x_{_{\chi_j^\pm}}\over x_{_{\rm w}}}\Big)^{1/2}
\Big[{\partial\varphi_1\over\partial x_{_{\chi_j^\pm}}}(x_{_{\chi_i^\pm}}
,x_{_{\chi_j^\pm}})
\nonumber\\
&&\hspace{1.2cm}
-{2-2x_{_{\chi_i^\pm}}\varrho_{_{0,1}}(x_{_{\chi_i^\pm}},x_{_{\chi_j^\pm}})
\over x_{_{\chi_i^\pm}}-x_{_{\chi_j^\pm}}}\cdot(1+\ln x_{_{\rm z}})\Big]
\Im\Big({\cal B}_{_{ji}}\xi^L_{_{ij}}+{\cal B}^\dagger_{_{ji}}
\xi^R_{_{ij}}\Big)\;.
\label{MD-z-G1}
\end{eqnarray}

\subsection{The effective Lagrangian from $\gamma Z$ sector}
\indent\indent
When a closed chargino loop is attached to the virtual $\gamma$ and $Z$ gauge bosons,
the corresponding correction to the effective Lagrangian is very tedious. If we ignore the terms
which are proportional to the suppression factor $1-4s_{_{\rm w}}^2$, the correction
from this sector to the effective Lagrangian is drastically simplified as
\begin{eqnarray}
&&{\cal L}_{_{\gamma Z}}=
{e^4\over8(4\pi)^2s_{_{\rm w}}^2c_{_{\rm w}}^2\Lambda^2}
\Big(\xi^L_{_{ii}}-\xi^R_{_{ii}}\Big)
\lim\limits_{x_{_{\chi_i^\pm}}\rightarrow x_{_{\chi_j^\pm}}}
T_3(x_{_{\rm z}},x_{_{\chi_i^\pm}},x_{_{\chi_j^\pm}})
\nonumber\\
&&\hspace{1.2cm}\times
\Big[\Big(T_{_f}^Z-Q_{_f}s_{_{\rm w}}^2\Big)({\cal O}_{_2}^-+{\cal O}_{_3}^-)
+Q_{_f}s_{_{\rm w}}^2({\cal O}_{_2}^++{\cal O}_{_3}^+)\Big]+\cdots\;.
\label{gamma-z}
\end{eqnarray}
Using the definitions of the matrices $\xi^{L,R}_{_{ij}}$ in Eq.(\ref{coupling-xi}),
one can find that the effective couplings $\xi^L_{_{ii}}-\xi^R_{_{ii}}\;(i=1,2)$ are
real. Correspondingly, the correction to the lepton MDMs from this sector is written as
\begin{eqnarray}
&&a_l^{\gamma Z}={e^4Q_{_f}m_{_l}^2\over4(4\pi)^4s_{_{\rm w}}^2c_{_{\rm w}}^2\Lambda^2}
\Big(\xi^L_{_{ii}}-\xi^R_{_{ii}}\Big)
\lim\limits_{x_{_{\chi_j^\pm}}\rightarrow x_{_{\chi_i^\pm}}}
T_3(x_{_{\rm z}},x_{_{\chi_i^\pm}},x_{_{\chi_j^\pm}})\;,
\label{MD-gamma-z}
\end{eqnarray}
and the correction to the lepton EDMs is zero. In the limit
$x_{_{\chi_i^\pm}}\gg x_{_{\rm z}}$, we can approximate the correction
to the lepton MDMs from this sector as
\begin{eqnarray}
&&a_l^{\gamma Z}={e^4Q_{_f}m_{_l}^2\over4(4\pi)^4s_{_{\rm w}}^2c_{_{\rm w}}^2\Lambda^2}
\Big(\xi^L_{_{ii}}-\xi^R_{_{ii}}\Big)\Big[
{13\over18x_{_{\chi_i^\pm}}}+{\ln x_{_{\chi_i^\pm}}-2\ln x_{_{\rm z}}
\over3x_{_{\chi_i^\pm}}}
\nonumber\\
&&\hspace{1.2cm}
+\lim\limits_{x_{_{\chi_j^\pm}}\rightarrow x_{_{\chi_i^\pm}}}
\Big(2x_{_{\chi_i^\pm}}{\partial^2\varphi_1\over\partial x_{_{\chi_i^\pm}}^2}
-{\partial\varphi_1\over\partial x_{_{\chi_i^\pm}}}\Big)
(x_{_{\chi_i^\pm}},x_{_{\chi_j^\pm}})\Big]\;.
\label{MD-gamma-z1}
\end{eqnarray}

\subsection{The effective Lagrangian from $W^\mp H^\pm\;(W^\mp G^\pm)$ sector}
\indent\indent
As a closed chargino-neutralino loop is attached to the virtual
$W^\pm$ gauge boson and charged Higgs $H^\mp$, the induced
Lagrangian can be written as
\begin{eqnarray}
&&{\cal L}_{_{WH}}=
-{e^4\tan\beta\over 16(4\pi)^2s_{_{\rm w}}^4c_{_{\rm w}}Q_{_f}\Lambda^2}
\Bigg\{\Big({x_{_{\chi_j^\pm}}\over x_{_{\rm w}}}\Big)^{1/2}
F_3(x_{_{\rm w}},x_{_{H^\pm}},x_{_{\chi_i^0}},x_{_{\chi_j^\pm}})
\Big[\Big(\sin\beta{\cal G}^{1L}_{_{ji}}\zeta^L_{_{ij}}
\nonumber\\
&&\hspace{1.2cm}
-\cos\beta{\cal G}^{1R}_{_{ji}}\zeta^R_{_{ij}}\Big){\cal O}_{_6}^-
+\Big(\sin\beta({\cal G}^{1L})^\dagger_{_{ij}}(\zeta^L)^\dagger_{_{ji}}
-\cos\beta({\cal G}^{1R})^\dagger_{_{ij}}(\zeta^R)^\dagger_{_{ji}}\Big){\cal O}_{_6}^+\Big]
\nonumber\\
&&\hspace{1.2cm}
+\Big({x_{_{\chi_i^0}}\over x_{_{\rm w}}}\Big)^{1/2}
F_4(x_{_{\rm w}},x_{_{H^\pm}},x_{_{\chi_i^0}},x_{_{\chi_j^\pm}})
\Big[\Big(\sin\beta{\cal G}^{1L}_{_{ji}}\zeta^R_{_{ij}}
-\cos\beta{\cal G}^{1R}_{_{ji}}\zeta^L_{_{ij}}\Big){\cal O}_{_6}^-
\nonumber\\
&&\hspace{1.2cm}
+\Big(\sin\beta({\cal G}^{1L})^\dagger_{_{ij}}(\zeta^R)^\dagger_{_{ji}}
-\cos\beta({\cal G}^{1R})^\dagger_{_{ij}}(\zeta^L)^\dagger_{_{ji}}\Big){\cal O}_{_6}^+\Big]
\nonumber\\
&&\hspace{1.2cm}
+\Big({x_{_{\chi_j^\pm}}\over x_{_{\rm w}}}\Big)^{1/2}
F_5(x_{_{\rm w}},x_{_{H^\pm}},x_{_{\chi_i^0}},x_{_{\chi_j^\pm}})
\Big[\Big(\sin\beta{\cal G}^{1L}_{_{ji}}\zeta^L_{_{ij}}
+\cos\beta{\cal G}^{1R}_{_{ji}}\zeta^R_{_{ij}}\Big){\cal O}_{_6}^-
\nonumber\\
&&\hspace{1.2cm}
+\Big(\sin\beta({\cal G}^{1L})^\dagger_{_{ij}}(\zeta^L)^\dagger_{_{ji}}
+\cos\beta({\cal G}^{1R})^\dagger_{_{ij}}(\zeta^R)^\dagger_{_{ji}}\Big){\cal O}_{_6}^+\Big]
\nonumber\\
&&\hspace{1.2cm}
+\Big({x_{_{\chi_i^0}}\over x_{_{\rm w}}}\Big)^{1/2}
F_6(x_{_{\rm w}},x_{_{H^\pm}},x_{_{\chi_i^0}},x_{_{\chi_j^\pm}})
\Big[\Big(\sin\beta{\cal G}^{1L}_{_{ji}}\zeta^R_{_{ij}}
+\cos\beta{\cal G}^{1R}_{_{ji}}\zeta^L_{_{ij}}\Big){\cal O}_{_6}^-
\nonumber\\
&&\hspace{1.2cm}
+\Big(\sin\beta({\cal G}^{1L})^\dagger_{_{ij}}(\zeta^R)^\dagger_{_{ji}}
+\cos\beta({\cal G}^{1R})^\dagger_{_{ij}}(\zeta^L)^\dagger_{_{ji}}\Big){\cal O}_{_6}^+\Big]\Bigg\}
\label{W-charged Higgs}
\end{eqnarray}
with
\begin{eqnarray}
&&\zeta^L_{ij}={\cal N}^\dagger_{i2}(U_{_L})_{_{1j}}
-{1\over\sqrt{2}}{\cal N}^\dagger_{i4}(U_{_L})_{_{2j}}
\;,\nonumber\\
&&\zeta^R_{ij}={\cal N}_{2i}(U_{_R}^\dagger)_{_{j1}}
+{1\over\sqrt{2}}{\cal N}_{3i}(U_{_R}^\dagger)_{_{j2}}
\;,\nonumber\\
&&{\cal G}^{1L}_{_{ji}}={1\over\sqrt{2}}(U_{_L})_{_{2j}}
\Big({\cal N}_{1i}s_{_{\rm w}}+{\cal N}_{2i}c_{_{\rm w}}\Big)
-(U_{_L})_{_{1j}}{\cal N}_{3i}c_{_{\rm w}}
\;,\nonumber\\
&&{\cal G}^{1R}_{_{ji}}={1\over\sqrt{2}}(U_{_R}^\dagger)_{_{j2}}
\Big({\cal N}^\dagger_{i1}s_{_{\rm w}}+{\cal N}^\dagger_{i2}c_{_{\rm w}}\Big)
-(U_{_R}^\dagger)_{_{j1}}{\cal N}^\dagger_{i4}c_{_{\rm w}}
\;,\nonumber\\
&&\;(i=1,\cdots,4,\;j=1,2)\;.
\label{coupling-zeta}
\end{eqnarray}
Here, the $4\times4$ matrix ${\cal N}$ denotes the mixing matrix of
the four neutralinos $\chi_i^0\;(i=1,\;\cdots,\;4)$.

The corresponding corrections to the lepton MDMs and EDMs are respectively expressed as
\begin{eqnarray}
&&a_l^{WH}=-{e^4m_{_l}^2\tan\beta\over4(4\pi)^4s_{_{\rm w}}^4c_{_{\rm w}}\Lambda^2}
\Bigg\{\Big({x_{_{\chi_j^\pm}}\over x_{_{\rm w}}}\Big)^{1/2}
F_3(x_{_{\rm w}},x_{_{H^\pm}},x_{_{\chi_i^0}},x_{_{\chi_j^\pm}})
\Re\Big(\sin\beta{\cal G}^{1L}_{_{ji}}\zeta^L_{_{ij}}
-\cos\beta{\cal G}^{1R}_{_{ji}}\zeta^R_{_{ij}}\Big)
\nonumber\\
&&\hspace{1.2cm}
+\Big({x_{_{\chi_i^0}}\over x_{_{\rm w}}}\Big)^{1/2}
F_4(x_{_{\rm w}},x_{_{H^\pm}},x_{_{\chi_i^0}},x_{_{\chi_j^\pm}})
\Re\Big(\sin\beta{\cal G}^{1L}_{_{ji}}\zeta^R_{_{ij}}
-\cos\beta{\cal G}^{1R}_{_{ji}}\zeta^L_{_{ij}}\Big)
\nonumber\\
&&\hspace{1.2cm}
+\Big({x_{_{\chi_j^\pm}}\over x_{_{\rm w}}}\Big)^{1/2}
F_5(x_{_{\rm w}},x_{_{H^\pm}},x_{_{\chi_i^0}},x_{_{\chi_j^\pm}})
\Re\Big(\sin\beta{\cal G}^{1L}_{_{ji}}\zeta^L_{_{ij}}
+\cos\beta{\cal G}^{1R}_{_{ji}}\zeta^R_{_{ij}}\Big)
\nonumber\\
&&\hspace{1.2cm}
+\Big({x_{_{\chi_i^0}}\over x_{_{\rm w}}}\Big)^{1/2}
F_6(x_{_{\rm w}},x_{_{H^\pm}},x_{_{\chi_i^0}},x_{_{\chi_j^\pm}})
\Re\Big(\sin\beta{\cal G}^{1L}_{_{ji}}\zeta^R_{_{ij}}
+\cos\beta{\cal G}^{1R}_{_{ji}}\zeta^L_{_{ij}}\Big)\Bigg\}\;,
\nonumber\\
&&d_l^{WH}=-{e^5m_{_l}\tan\beta\over8(4\pi)^4s_{_{\rm w}}^4c_{_{\rm w}}\Lambda^2}
\Bigg\{\Big({x_{_{\chi_j^\pm}}\over x_{_{\rm w}}}\Big)^{1/2}
F_3(x_{_{\rm w}},x_{_{H^\pm}},x_{_{\chi_i^0}},x_{_{\chi_j^\pm}})
\Im\Big(\sin\beta{\cal G}^{1L}_{_{ji}}\zeta^L_{_{ij}}
-\cos\beta{\cal G}^{1R}_{_{ji}}\zeta^R_{_{ij}}\Big)
\nonumber\\
&&\hspace{1.2cm}
+\Big({x_{_{\chi_i^0}}\over x_{_{\rm w}}}\Big)^{1/2}
F_4(x_{_{\rm w}},x_{_{H^\pm}},x_{_{\chi_i^0}},x_{_{\chi_j^\pm}})
\Im\Big(\sin\beta{\cal G}^{1L}_{_{ji}}\zeta^R_{_{ij}}
-\cos\beta{\cal G}^{1R}_{_{ji}}\zeta^L_{_{ij}}\Big)
\nonumber\\
&&\hspace{1.2cm}
+\Big({x_{_{\chi_j^\pm}}\over x_{_{\rm w}}}\Big)^{1/2}
F_5(x_{_{\rm w}},x_{_{H^\pm}},x_{_{\chi_i^0}},x_{_{\chi_j^\pm}})
\Im\Big(\sin\beta{\cal G}^{1L}_{_{ji}}\zeta^L_{_{ij}}
+\cos\beta{\cal G}^{1R}_{_{ji}}\zeta^R_{_{ij}}\Big)
\nonumber\\
&&\hspace{1.2cm}
+\Big({x_{_{\chi_i^0}}\over x_{_{\rm w}}}\Big)^{1/2}
F_6(x_{_{\rm w}},x_{_{H^\pm}},x_{_{\chi_i^0}},x_{_{\chi_j^\pm}})
\Im\Big(\sin\beta{\cal G}^{1L}_{_{ji}}\zeta^R_{_{ij}}
+\cos\beta{\cal G}^{1R}_{_{ji}}\zeta^L_{_{ij}}\Big)\Bigg\}\;,
\label{MED-W-H}
\end{eqnarray}
where the concrete expressions of $F_{3,4,5,6}$ can be found in appendix.
The corrections from this sector to the MDM of lepton
are decided by a linear combination of real parts of the effective couplings
$\sin\beta{\cal G}^{1L}_{_{ji}}\zeta^L_{_{ij}}-\cos\beta{\cal G}^{1R}_{_{ji}}\zeta^R_{_{ij}}$,
$\sin\beta{\cal G}^{1L}_{_{ji}}\zeta^R_{_{ij}}-\cos\beta{\cal G}^{1R}_{_{ji}}\zeta^L_{_{ij}}$,
$\sin\beta{\cal G}^{1L}_{_{ji}}\zeta^L_{_{ij}}+\cos\beta{\cal G}^{1R}_{_{ji}}\zeta^R_{_{ij}}$,
as well as $\sin\beta{\cal G}^{1L}_{_{ji}}\zeta^R_{_{ij}}+\cos\beta{\cal G}^{1R}_{_{ji}}\zeta^L_{_{ij}}$,
and the corrections from this sector to the EDM of lepton
are decided by a linear combination of imaginary parts of those effective couplings.
Using the asymptotic expansion of the two-loop vacuum integral $\Phi(x,y,z)$
presented in appendix, we can simplify the expressions of Eq.\ref{MED-W-H}
in the limit $x_{_{\chi_i^0}},x_{_{\chi_j^\pm}}\gg x_{_{\rm w}}$.

As a closed chargino-neutralino loop is attached to the virtual
$W^\pm$ gauge boson and charged Goldstone $G^\mp$, the corresponding corrections
to the lepton MDMs and EDMs are similarly formulated as
\begin{eqnarray}
&&a_l^{WG}=-{e^4m_{_l}^2\over4(4\pi)^4s_{_{\rm w}}^4c_{_{\rm w}}\Lambda^2}
\Bigg\{\Big({x_{_{\chi_j^\pm}}\over x_{_{\rm w}}}\Big)^{1/2}
F_3(x_{_{\rm w}},x_{_{\rm w}},x_{_{\chi_i^0}},x_{_{\chi_j^\pm}})
\Re\Big(\cos\beta{\cal G}^{1L}_{_{ji}}\zeta^L_{_{ij}}
+\sin\beta{\cal G}^{1R}_{_{ji}}\zeta^R_{_{ij}}\Big)
\nonumber\\
&&\hspace{1.2cm}
+\Big({x_{_{\chi_i^0}}\over x_{_{\rm w}}}\Big)^{1/2}
F_4(x_{_{\rm w}},x_{_{\rm w}},x_{_{\chi_i^0}},x_{_{\chi_j^\pm}})
\Re\Big(\cos\beta{\cal G}^{1L}_{_{ji}}\zeta^R_{_{ij}}
+\sin\beta{\cal G}^{1R}_{_{ji}}\zeta^L_{_{ij}}\Big)
\nonumber\\
&&\hspace{1.2cm}
+\Big({x_{_{\chi_j^\pm}}\over x_{_{\rm w}}}\Big)^{1/2}
F_5(x_{_{\rm w}},x_{_{\rm w}},x_{_{\chi_i^0}},x_{_{\chi_j^\pm}})
\Re\Big(\cos\beta{\cal G}^{1L}_{_{ji}}\zeta^L_{_{ij}}
-\sin\beta{\cal G}^{1R}_{_{ji}}\zeta^R_{_{ij}}\Big)
\nonumber\\
&&\hspace{1.2cm}
+\Big({x_{_{\chi_i^0}}\over x_{_{\rm w}}}\Big)^{1/2}
F_6(x_{_{\rm w}},x_{_{\rm w}},x_{_{\chi_i^0}},x_{_{\chi_j^\pm}})
\Re\Big(\cos\beta{\cal G}^{1L}_{_{ji}}\zeta^R_{_{ij}}
-\sin\beta{\cal G}^{1R}_{_{ji}}\zeta^L_{_{ij}}\Big)\Bigg\}\;,
\nonumber\\
&&d_l^{WG}=-{e^5m_{_l}\over8(4\pi)^4s_{_{\rm w}}^4c_{_{\rm w}}\Lambda^2}
\Bigg\{\Big({x_{_{\chi_j^\pm}}\over x_{_{\rm w}}}\Big)^{1/2}
F_3(x_{_{\rm w}},x_{_{\rm w}},x_{_{\chi_i^0}},x_{_{\chi_j^\pm}})
\Im\Big(\cos\beta{\cal G}^{1L}_{_{ji}}\zeta^L_{_{ij}}
+\sin\beta{\cal G}^{1R}_{_{ji}}\zeta^R_{_{ij}}\Big)
\nonumber\\
&&\hspace{1.2cm}
+\Big({x_{_{\chi_i^0}}\over x_{_{\rm w}}}\Big)^{1/2}
F_4(x_{_{\rm w}},x_{_{\rm w}},x_{_{\chi_i^0}},x_{_{\chi_j^\pm}})
\Im\Big(\cos\beta{\cal G}^{1L}_{_{ji}}\zeta^R_{_{ij}}
+\sin\beta{\cal G}^{1R}_{_{ji}}\zeta^L_{_{ij}}\Big)
\nonumber\\
&&\hspace{1.2cm}
+\Big({x_{_{\chi_j^\pm}}\over x_{_{\rm w}}}\Big)^{1/2}
F_5(x_{_{\rm w}},x_{_{\rm w}},x_{_{\chi_i^0}},x_{_{\chi_j^\pm}})
\Im\Big(\cos\beta{\cal G}^{1L}_{_{ji}}\zeta^L_{_{ij}}
-\sin\beta{\cal G}^{1R}_{_{ji}}\zeta^R_{_{ij}}\Big)
\nonumber\\
&&\hspace{1.2cm}
+\Big({x_{_{\chi_i^0}}\over x_{_{\rm w}}}\Big)^{1/2}
F_6(x_{_{\rm w}},x_{_{\rm w}},x_{_{\chi_i^0}},x_{_{\chi_j^\pm}})
\Im\Big(\cos\beta{\cal G}^{1L}_{_{ji}}\zeta^R_{_{ij}}
-\sin\beta{\cal G}^{1R}_{_{ji}}\zeta^L_{_{ij}}\Big)\Bigg\}\;.
\label{MED-W-G}
\end{eqnarray}
Similarly, the corrections from this sector to the MDM of lepton
depend on a linear combination of real parts of the effective couplings
$\cos\beta{\cal G}^{1L}_{_{ji}}\zeta^L_{_{ij}}+\sin\beta{\cal G}^{1R}_{_{ji}}\zeta^R_{_{ij}}$,
$\cos\beta{\cal G}^{1L}_{_{ji}}\zeta^R_{_{ij}}+\sin\beta{\cal G}^{1R}_{_{ji}}\zeta^L_{_{ij}}$,
$\cos\beta{\cal G}^{1L}_{_{ji}}\zeta^L_{_{ij}}-\sin\beta{\cal G}^{1R}_{_{ji}}\zeta^R_{_{ij}}$,
as well as $\cos\beta{\cal G}^{1L}_{_{ji}}\zeta^R_{_{ij}}-\sin\beta{\cal G}^{1R}_{_{ji}}\zeta^L_{_{ij}}$,
and the corrections from this sector to the EDM of lepton
depend on a linear combination of imaginary parts of those effective couplings.

The contributions from those above sectors to effective Lagrangian
do not contain ultraviolet divergence. In the pieces discussed below,
the coefficients of high dimensional operators in effective Lagrangian
contain ultraviolet divergence that is caused by the divergent
subdiagrams. In order to obtain physical predictions of lepton
MDMs and EDMs, it is necessary to adopt a concrete renormalization
scheme removing the ultraviolet divergence. In literature, the on-shell renormalization
scheme is adopted frequently to subtract the ultraviolet
divergence which appears in the radiative electroweak corrections
\cite{onshell}. As an over-subtract scheme, the counter terms
include some finite terms which originate from those renormalization conditions
in the on-shell scheme beside the ultraviolet divergence to cancel
the corresponding ultraviolet divergence contained by the bare Lagrangian. In the concrete
calculation performed here, we apply this scheme to subtract the ultraviolet divergence
caused by the divergent subdiagrams.

\subsection{The effective Lagrangian from the $ZZ$ sector}
\indent\indent
The self energy of $Z$ gauge boson composed of a closed chargino loop
induces the ultraviolet divergence in the Wilson coefficients of effective
Lagrangian. Generally, the unrenormalized self energy of the weak gauge boson
$Z$ can be written as
\begin{eqnarray}
&&\Sigma_{_{\mu\nu}}^{\rm Z}(p)=\Lambda^2A_0^zg_{\mu\nu}+\Big(A_1^z+{p^2\over\Lambda^2}A_2^z\Big)
(p^2g_{\mu\nu}-p_\mu p_\nu)+\Big(B_1^z+{p^2\over\Lambda^2}B_2^z\Big)p_\mu p_\nu\;.
\label{eq-z1}
\end{eqnarray}
Correspondingly, the counter terms are given as
\begin{eqnarray}
&&\Sigma_{_{\mu\nu}}^{\rm ZC}(p)=-(\delta m_{_{\rm z}}^2+m_{_{\rm z}}^2\delta Z_{_{\rm z}})g_{\mu\nu}
-\delta Z_{_{\rm z}}(p^2g_{\mu\nu}-p_\mu p_\nu)\;.
\label{eq-z2}
\end{eqnarray}
The renormalized self energy is given by
\begin{eqnarray}
&&\hat{\Sigma}_{_{\mu\nu}}^{\rm Z}(p)=\Sigma_{_{\mu\nu}}^{\rm Z}(p)
+\Sigma_{_{\mu\nu}}^{\rm ZC}(p)\;.
\label{eq-z3}
\end{eqnarray}
For on-shell external gauge boson $Z$, we have \cite{onshell}
\begin{eqnarray}
&&\hat{\Sigma}_{_{\mu\nu}}^{\rm Z}(p)\epsilon^\nu(p)\Big|_{p^2=m_{_{\rm z}}^2}=0
\;,\nonumber\\
&&\lim\limits_{p^2\rightarrow m_{_{\rm z}}^2}{1\over p^2-m_{_{\rm z}}^2}
\hat{\Sigma}_{_{\mu\nu}}^{\rm Z}(p)\epsilon^\nu(p)=\epsilon_{_\mu}(p)\;,
\label{eq-z4}
\end{eqnarray}
where $\epsilon(p)$ is the polarization vector of $Z$ gauge boson.
From Eq. (\ref{eq-z4}), we get the counter terms
\begin{eqnarray}
&&\delta Z_{_{\rm z}}=A_1^z+{m_{_{\rm z}}^2\over\Lambda^2}A_2^z=A_1^z+x_{_{\rm z}}A_2^z\;,
\nonumber\\
&&\delta m_{_{\rm z}}^2=A_0^z\Lambda^2-m_{_{\rm z}}^2\delta Z_{_{\rm z}}\;.
\label{eq-z5}
\end{eqnarray}

\begin{figure}[t]
\setlength{\unitlength}{1mm}
\begin{center}
\begin{picture}(0,40)(0,0)
\put(-60,-100){\includegraphics{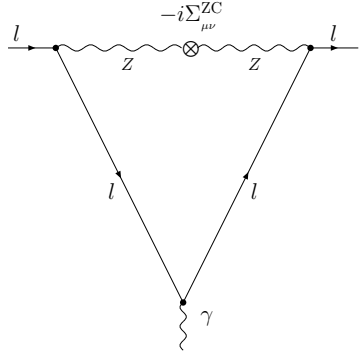}}
\end{picture}
\caption[]{The counter term diagram to cancel the ultraviolet caused
by the self energy of $Z$ boson.}
\label{fig2}
\end{center}
\end{figure}

Accordingly, the effective Lagrangian originating from the counter term
diagram (Fig.\ref{fig2}) can be formulated as
\begin{eqnarray}
&&\delta{\cal L}_{_{ZZ}}^{C}=
-{e^4\over12(4\pi)^2s_{_{\rm w}}^4c_{_{\rm w}}^4\Lambda^2}(4\pi x_{_{\rm R}})^{2\varepsilon}
{\Gamma^2(1+\varepsilon)\over(1-\varepsilon)^2}\Bigg\{\Big(\xi^L_{_{ji}}\xi^L_{_{ij}}
+\xi^R_{_{ji}}\xi^R_{_{ij}}\Big)\Big[
-{1\over\varepsilon}{x_{_{\chi_i^\pm}}+x_{_{\chi_j^\pm}}\over x_{_{\rm z}}^2}
\nonumber\\
&&\hspace{1.6cm}
+{5(x_{_{\chi_i^\pm}}+x_{_{\chi_j^\pm}})\over12x_{_{\rm z}}^2}
+{\varrho_{_{2,1}}(x_{_{\chi_i^\pm}},x_{_{\chi_j^\pm}})\over x_{_{\rm z}}^2}
+{5\over12x_{_{\rm z}}}+{x_{_{\chi_i^\pm}}+x_{_{\chi_j^\pm}}\over x_{_{\rm z}}^2}
\ln x_{_{\rm R}}\Big]
\nonumber\\
&&\hspace{1.6cm}
+2(x_{_{\chi_i^\pm}}x_{_{\chi_j^\pm}})^{1/2}\Big(\xi^L_{_{ji}}
\xi^R_{_{ij}}+\xi^R_{_{ji}}\xi^L_{_{ij}}\Big)
\Big[{1\over\varepsilon x_{_{\rm z}}^2}
-{\varrho_{_{1,1}}(x_{_{\chi_i^\pm}},x_{_{\chi_j^\pm}})\over x_{_{\rm z}}^2}
+{1\over12x_{_{\rm z}}^2}-{\ln x_{_{\rm R}}\over x_{_{\rm z}}^2}\Big]\Bigg\}
\nonumber\\
&&\hspace{1.6cm}\times
\Big[\Big(T_{_f}^Z-Q_{_f}s_{_{\rm w}}^2\Big)^2({\cal O}_{_2}^-+{\cal O}_{_3}^-)
+Q_{_f}^2s_{_{\rm w}}^4({\cal O}_{_2}^++{\cal O}_{_3}^+)\Big]
\nonumber\\
&&\hspace{1.6cm}
+{e^4\over4(4\pi)^2s_{_{\rm w}}^4c_{_{\rm w}}^4\Lambda^2}(4\pi x_{_{\rm R}})^{2\varepsilon}
{\Gamma^2(1+\varepsilon)\over(1-\varepsilon)^2}\Bigg\{\Big(\xi^L_{_{ji}}\xi^L_{_{ij}}
+\xi^R_{_{ji}}\xi^R_{_{ij}}\Big)\Big[
{1\over\varepsilon}{x_{_{\chi_i^\pm}}+x_{_{\chi_j^\pm}}\over x_{_{\rm z}}^2}
\nonumber\\
&&\hspace{1.6cm}
-{\varrho_{_{2,1}}(x_{_{\chi_i^\pm}},x_{_{\chi_j^\pm}})\over x_{_{\rm z}}^2}
-{x_{_{\chi_i^\pm}}+x_{_{\chi_j^\pm}}\over x_{_{\rm z}}^2}({7\over2}+\ln x_{_l}-\ln x_{_{\rm z}})
+{1\over4x_{_{\rm z}}}-{x_{_{\chi_i^\pm}}+x_{_{\chi_j^\pm}}\over x_{_{\rm z}}^2}
\ln x_{_{\rm R}}\Big]
\nonumber\\
&&\hspace{1.6cm}
+2(x_{_{\chi_i^\pm}}x_{_{\chi_j^\pm}})^{1/2}\Big(\xi^L_{_{ji}}
\xi^R_{_{ij}}+\xi^R_{_{ji}}\xi^L_{_{ij}}\Big)
\Big[-{1\over\varepsilon x_{_{\rm z}}^2}
+{\varrho_{_{1,1}}(x_{_{\chi_i^\pm}},x_{_{\chi_j^\pm}})\over x_{_{\rm z}}^2}
\nonumber\\
&&\hspace{1.6cm}
+{1\over x_{_{\rm z}}^2}(3+\ln x_{_l}-\ln x_{_{\rm z}})
+{\ln x_{_{\rm R}}\over x_{_{\rm z}}^2}\Big]\Bigg\}
Q_{_f}s_{_{\rm w}}^2\Big(T_{_f}^Z-Q_{_l}s_{_{\rm w}}^2\Big)
({\cal O}_{_6}^-+{\cal O}_{_6}^+)+\cdots\;.
\label{counter-eff}
\end{eqnarray}
Here, $\varepsilon=2-D/2$ with $D$ representing the time-space dimension,
and $x_{_{\rm R}}=\Lambda_{_{\rm RE}}^2/\Lambda^2$ ($\Lambda_{_{\rm RE}}$
denotes the renormalization scale).

As a result of the preparation mentioned above, we can add the contributions from
the counter term diagram to cancel the corresponding ultraviolet
divergence contained by the bare effective Lagrangian.
Using the definitions of the matrices $\xi^{L,R}_{_{ij}}$ in Eq.(\ref{coupling-xi}),
we derive $\xi^{L}_{_{ij}}=\xi^{L*}_{_{ji}},\;\xi^{R}_{_{ij}}=\xi^{R*}_{_{ji}}$.
The resulted theoretical predictions on the lepton MDMs and EDMs are respectively
written as
\begin{eqnarray}
&&a_{l,\chi^\pm}^{ZZ}=
-{e^4m_{_l}^2\over(4\pi)^4s_{_{\rm w}}^4c_{_{\rm w}}^4\Lambda^2}\Bigg\{
\Big(|\xi^L_{_{ij}}|^2+|\xi^R_{_{ij}}|^2\Big)
\Big[\Big(T_{_f}^Z-Q_{_f}s_{_{\rm w}}^2\Big)^2+Q_{_f}^2s_{_{\rm w}}^4\Big]
\nonumber\\
&&\hspace{1.2cm}\times
\Bigg[{Q_{_f}\over3}\Big(T_5(x_{_{\rm z}},x_{_{\chi_i^\pm}},x_{_{\chi_j^\pm}})
+{x_{_{\chi_i^\pm}}+x_{_{\chi_j^\pm}}\over x_{_{\rm z}}^2}\ln x_{_{\rm R}}\Big)
+{1\over4}T_4(x_{_{\rm z}},x_{_{\chi_i^\pm}},x_{_{\chi_j^\pm}})\Bigg]
\nonumber\\
&&\hspace{1.2cm}
+{1\over8}\Big(|\xi^L_{_{ij}}|^2-|\xi^R_{_{ij}}|^2\Big)
\Big[\Big(T_{_f}^Z-Q_{_f}s_{_{\rm w}}^2\Big)^2-Q_{_f}^2s_{_{\rm w}}^4\Big]
T_{6}(x_{_{\rm z}},x_{_{\chi_i^\pm}},x_{_{\chi_j^\pm}})
\nonumber\\
&&\hspace{1.2cm}
-\Re(\xi^L_{_{ij}}\xi^R_{_{ji}})
\Big[\Big(T_{_f}^Z-Q_{_f}s_{_{\rm w}}^2\Big)^2+Q_{_f}^2s_{_{\rm w}}^4\Big]
(x_{_{\chi_i^\pm}}x_{_{\chi_j^\pm}})^{1/2}
\nonumber\\
&&\hspace{1.2cm}\times
\Bigg[{1\over4}T_{7}(x_{_{\rm z}},x_{_{\chi_i^\pm}},x_{_{\chi_j^\pm}})
+{4Q_{_f}\over3x_{_{\rm z}}^2}\ln{x_{_{\rm z}}\over x_{_{\rm R}}}-{7Q_{_f}\over3x_{_{\rm z}}^2}\Bigg]
\nonumber\\
&&\hspace{1.2cm}
-\Big(|\xi^L_{_{ij}}|^2+|\xi^R_{_{ij}}|^2\Big)s_{_{\rm w}}^2
\Big(T_{_f}^Z-Q_{_f}s_{_{\rm w}}^2\Big)
\Bigg[{Q_{_f}\over4}T_{9}(x_{_{\rm z}},x_{_{\chi_i^\pm}},x_{_{\chi_j^\pm}})
\nonumber\\
&&\hspace{1.2cm}
-{Q_{_f}^2\over4x_{_{\rm z}}}+{Q_{_f}^2\over x_{_{\rm z}}^2}(2-\ln{x_{_{\rm z}}
\over x_{_{\rm R}}})(x_{_{\chi_i^\pm}}+x_{_{\chi_j^\pm}})
-{Q_{_f}^2\over2x_{_{\rm z}}^2}(x_{_{\chi_i^\pm}}\ln x_{_{\chi_i^\pm}}
+x_{_{\chi_j^\pm}}\ln x_{_{\chi_j^\pm}})
\nonumber\\
&&\hspace{1.2cm}
+{Q_{_f}^2\over2x_{_{\rm z}}^2}\cdot(\varrho_{_{2,1}}(x_{_{\chi_i^\pm}},x_{_{\chi_j^\pm}})
-x_{_{\chi_i^\pm}}x_{_{\chi_j^\pm}}\varrho_{_{0,1}}(x_{_{\chi_i^\pm}},x_{_{\chi_j^\pm}}))\Bigg]
\nonumber\\
&&\hspace{1.2cm}
-4Q_{_f}^2\Re(\xi^L_{_{ij}}\xi^R_{_{ji}})
s_{_{\rm w}}^2\Big(T_{_f}^Z-Q_{_f}s_{_{\rm w}}^2\Big)
(x_{_{\chi_i^\pm}}x_{_{\chi_j^\pm}})^{1/2}
{2-\ln x_{_{\rm z}}+\ln x_{_{\rm R}}\over x_{_{\rm z}}^2}\Bigg\}\;,
\nonumber\\
&&d_{l,\chi^\pm}^{ZZ}=
{e^5m_{_l}\over(4\pi)^4s_{_{\rm w}}^4c_{_{\rm w}}^4\Lambda^2}\cdot
\Im(\xi^L_{_{ij}}\xi^R_{_{ji}})
(x_{_{\chi_i^\pm}}x_{_{\chi_j^\pm}})^{1/2}\Bigg\{
Q_{_f}s_{_{\rm w}}^2\Big(T_{_f}^Z-Q_{_f}s_{_{\rm w}}^2\Big)
\nonumber\\
&&\hspace{1.2cm}\times
\Big({\partial^2\over\partial x_{_{\rm z}}\partial x_{_{\chi_j^\pm}}}
-{\partial^2\over\partial x_{_{\rm z}}\partial x_{_{\chi_i^\pm}}}\Big)
\Big({\Phi(x_{_{\rm z}},x_{_{\chi_i^\pm}},x_{_{\chi_j^\pm}})
-\varphi_0(x_{_{\chi_i^\pm}},x_{_{\chi_j^\pm}})
\over x_{_{\rm z}}}\Big)
\nonumber\\
&&\hspace{1.2cm}
-{1\over16}\Big[\Big(T_{_f}^Z-Q_{_f}s_{_{\rm w}}^2\Big)^2+Q_{_f}^2s_{_{\rm w}}^4\Big]
T_{8}(x_{_{\rm z}},x_{_{\chi_i^\pm}},x_{_{\chi_j^\pm}})\Bigg\}\;.
\label{zz-chargino}
\end{eqnarray}
In other words, the corrections from this sector to the MDM of lepton
are decided by a linear combination of the real effective couplings
$|\xi^L_{_{ij}}|^2\pm|\xi^R_{_{ij}}|^2$ and real parts of the effective
couplings $\xi^L_{_{ij}}\xi^R_{_{ji}}$, and the corrections from this sector
to the EDM of lepton are proportional to imaginary parts of the effective
couplings $\xi^L_{_{ij}}\xi^R_{_{ji}}$.

Because a real photon can not be attached to the internal closed neutralino loop,
the corresponding effective Lagrangian only contains the corrections to the
lepton MDMs:
\begin{eqnarray}
&&a_{l,\chi^0}^{ZZ}=
-{e^4Q_{_f}m_{_l}^2\over(4\pi)^4s_{_{\rm w}}^4c_{_{\rm w}}^4\Lambda^2}
\Bigg\{-{1\over3}\Big(|\eta^L_{_{ij}}|^2+|\eta^R_{_{ij}}|^2\Big)\Big[
\Big(T_{_f}^Z-Q_{_f}s_{_{\rm w}}^2\Big)^2
+Q_{_f}^2s_{_{\rm w}}^4\Big]
\nonumber\\
&&\hspace{1.2cm}\times
\Big(T_5(x_{_{\rm z}},x_{_{\chi_i^0}},x_{_{\chi_j^0}})
+{x_{_{\chi_i^0}}+x_{_{\chi_j^0}}\over x_{_{\rm z}}^2}\ln x_{_{\rm R}}\Big)
\nonumber\\
&&\hspace{1.2cm}
+{1\over3}\Re(\eta^L_{_{ij}}\eta^R_{_{ji}})
\Big[\Big(T_{_f}^Z-Q_{_f}s_{_{\rm w}}^2\Big)^2+Q_{_f}^2s_{_{\rm w}}^4\Big]
(x_{_{\chi_i^0}}x_{_{\chi_j^0}})^{1/2}
\Bigg[{4\over x_{_{\rm z}}^2}\ln{x_{_{\rm z}}\over x_{_{\rm R}}}-{7\over x_{_{\rm z}}^2}\Bigg]
\nonumber\\
&&\hspace{1.2cm}
+{1\over2 x_{_{\rm z}}^2}\Big(|\eta^L_{_{ij}}|^2+|\eta^R_{_{ij}}|^2\Big)Q_{_f}s_{_{\rm w}}^2
\Big(T_{_f}^Z-Q_{_f}s_{_{\rm w}}^2\Big)\Bigg[{x_{_{\rm z}}\over2}
+(x_{_{\chi_i^0}}\ln x_{_{\chi_i^0}}+x_{_{\chi_j^0}}\ln x_{_{\chi_j^0}})
\nonumber\\
&&\hspace{1.2cm}
-2(x_{_{\chi_i^0}}+x_{_{\chi_j^0}})(2-\ln{x_{_{\rm z}}\over x_{_{\rm R}}})
-\varrho_{_{2,1}}(x_{_{\chi_i^0}},x_{_{\chi_j^0}})
+x_{_{\chi_i^0}}x_{_{\chi_j^0}}\varrho_{_{0,1}}
(x_{_{\chi_i^0}},x_{_{\chi_j^0}})\Bigg]
\nonumber\\
&&\hspace{1.2cm}
-4Q_{_f}\Re(\eta^L_{_{ij}}\eta^R_{_{ji}})
s_{_{\rm w}}^2\Big(T_{_f}^Z-Q_{_f}s_{_{\rm w}}^2\Big)
(x_{_{\chi_i^0}}x_{_{\chi_j^0}})^{1/2}
{2-\ln x_{_{\rm z}}+\ln x_{_{\rm R}}\over x_{_{\rm z}}^2}\Bigg\}
\label{zz-neutralino}
\end{eqnarray}
with
\begin{eqnarray}
&&\eta^L_{ij}={\cal N}_{i4}^\dagger{\cal N}_{4j}
\;,\nonumber\\
&&\eta^R_{ij}={\cal N}_{j3}^\dagger{\cal N}_{3i}
\;,(i,j=1,\cdots,4)\;.
\label{coupling-eta}
\end{eqnarray}
In order to get Eq.(\ref{zz-neutralino}), we apply unitary property
of the matrices $\eta^{L,R}$. The corrections from this sector to the MDM of lepton
depend on a linear combination of the real effective couplings
$|\eta^L_{_{ij}}|^2\pm|\eta^R_{_{ij}}|^2$ and real parts of the effective
couplings $\eta^L_{_{ij}}\eta^R_{_{ji}}$, and the corrections from this sector
to the EDM of lepton are proportional to imaginary parts of the effective
couplings $\eta^L_{_{ij}}\eta^R_{_{ji}}$.

We can also simplify Eq.(\ref{zz-chargino}) and Eq.(\ref{zz-neutralino}) in the limit $
x_{_{\chi_i^\pm}},\;x_{_{\chi_j^\pm}},\;x_{_{\chi_i^0}},\;x_{_{\chi_j^0}}
\gg x_{_{\rm z}}$ using the asymptotic expansion of $\Phi(x,y,z)$.
The concrete expressions of $T_4\sim T_{9}$ can be found in appendix.

\subsection{The effective Lagrangian from the $WW$ sector}
\indent\indent
Similarly, the self energy of $W$ gauge boson composed of a closed chargino-neutralino loop
induces the ultraviolet divergence in the Wilson coefficients of effective
Lagrangian. Accordingly, the unrenormalized $W$ self energy is expressed as
\begin{eqnarray}
&&\Sigma_{_{\mu\nu}}^{\rm W}(p)=\Lambda^2A_0^{\rm w}g_{\mu\nu}+\Big(A_1^{\rm w}
+{p^2\over\Lambda^2}A_2^{\rm w}\Big)(p^2g_{\mu\nu}-p_\mu p_\nu)
+\Big(B_1^{\rm w}+{p^2\over\Lambda^2}B_2^{\rm w}\Big)p_\mu p_\nu\;.
\label{eq-w1}
\end{eqnarray}
The corresponding counter terms are given as
\begin{eqnarray}
&&\Sigma_{_{\mu\nu}}^{\rm WC}(p)=-(\delta m_{_{\rm w}}^2+m_{_{\rm w}}^2\delta Z_{_{\rm w}})g_{\mu\nu}
-\delta Z_{_{\rm w}}(p^2g_{\mu\nu}-p_\mu p_\nu)\;.
\label{eq-w2}
\end{eqnarray}

\begin{figure}[t]
\setlength{\unitlength}{1mm}
\begin{center}
\begin{picture}(0,100)(0,0)
\put(-60,-40){\includegraphics{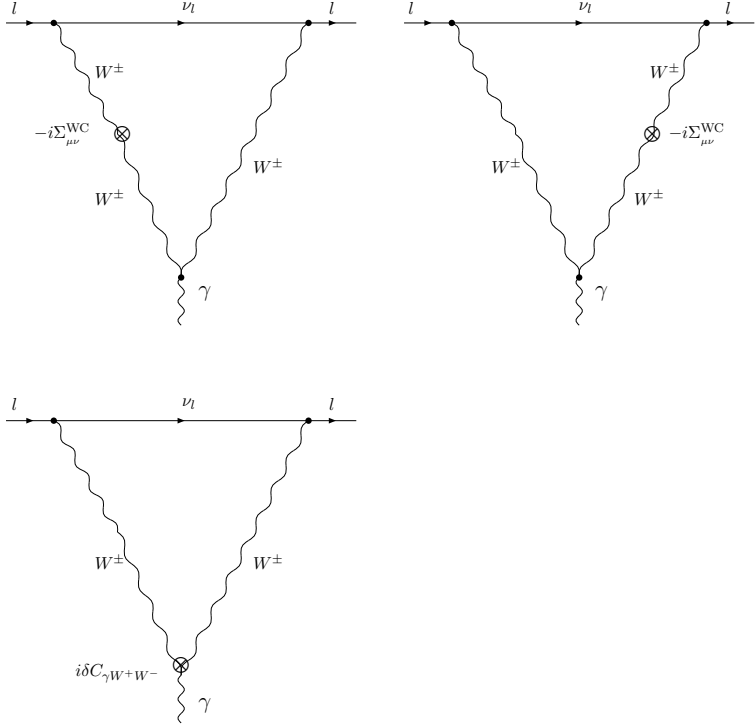}}
\end{picture}
\caption[]{The counter term diagram to cancel the ultraviolet caused
by the self energy of $W$ boson and electroweak radiative corrections to
$\gamma W^+W^-$ vertex.}
\label{fig3}
\end{center}
\end{figure}

The renormalized self energy is given by
\begin{eqnarray}
&&\hat{\Sigma}_{_{\mu\nu}}^{\rm W}(p)=\Sigma_{_{\mu\nu}}^{\rm W}(p)+\Sigma_{_{\mu\nu}}^{\rm WC}(p)
\label{eq-w3}
\end{eqnarray}
For on-shell external gauge boson $W^\pm$, we have \cite{onshell}
\begin{eqnarray}
&&\hat{\Sigma}_{_{\mu\nu}}^{\rm W}(p)\epsilon^\nu(p)\Big|_{p^2=m_{_{\rm w}}^2}=0
\;,\nonumber\\
&&\lim\limits_{p^2\rightarrow m_{_{\rm w}}^2}{1\over p^2-m_{_{\rm w}}^2}
\hat{\Sigma}_{_{\mu\nu}}^{\rm W}(p)\epsilon^\nu(p)=\epsilon_{_\mu}(p)\;,
\label{eq-w4}
\end{eqnarray}
where $\epsilon(p)$ is the polarization vector of $W$ gauge boson.
Inserting Eq. (\ref{eq-w1}) and Eq. (\ref{eq-w2}) into Eq. (\ref{eq-w4}),
we derive the counter terms for the $W$ self energy as
\begin{eqnarray}
&&\delta Z_{_{\rm w}}=A_1^{\rm w}+{m_{_{\rm w}}^2\over\Lambda^2}A_2^{\rm w}
=A_1^{\rm w}+x_{_{\rm z}}A_2^{\rm w}\;,
\nonumber\\
&&\delta m_{_{\rm w}}^2=A_0^{\rm w}\Lambda^2-m_{_{\rm w}}^2\delta Z_{_{\rm w}}\;.
\label{eq-w5}
\end{eqnarray}
Differing from the analysis in the $ZZ$ sector, we should derive the
counter term for the vertex $\gamma W^+W^-$ here since the corresponding coupling
is not zero at tree level. In the nonlinear $R_\xi$ gauge with $\xi=1$,
the counter term for the vertex $\gamma W^+W^-$ is
\begin{eqnarray}
&&i\delta C_{\gamma W^+W^-}=ie\cdot\delta Z_{_{\rm w}}\Big[g_{\mu\nu}(k_1-k_2)_\rho
+g_{\nu\rho}(k_2-k_3)_\mu+g_{\rho\mu}(k_3-k_1)_\nu\Big]\;,
\label{eq-w6}
\end{eqnarray}
where $k_i\;(i=1,\;2,\;3)$ denote the injection momenta of $W^\pm$ and photon,
and $\mu,\;\nu,\;\rho$ denote the corresponding Lorentz indices respectively.

We present the counter term diagrams to cancel the ultraviolet divergence contained in
the bare effective Lagrangian from $WW$
sector in Fig.\ref{fig3}, and we can verify that the sum of corresponding amplitude
satisfies the Ward identity required by the QED gauge invariance obviously.
Accordingly, the effective Lagrangian originating from the counter
term diagrams can be written as

\begin{eqnarray}
&&\delta{\cal L}_{_{WW}}^C=
{e^4\over(4\pi)^2s_{_{\rm w}}^4\Lambda^2Q_{_f}}(4\pi x_{_{\rm R}})^{2\varepsilon}
{\Gamma^2(1+\varepsilon)\over(1-\varepsilon)^2}\Bigg\{\Big(\zeta^{L*}_{_{ij}}
\zeta^L_{_{ij}}+\zeta^{R*}_{_{ij}}\zeta^R_{_{ij}}\Big)
\nonumber\\
&&\hspace{1.4cm}\times
\Big[{5\over24x_{_{\rm w}}^2}\Big(-{x_{_{\chi_i^0}}+x_{_{\chi_j^\pm}}\over\varepsilon}
-{x_{_{\chi_i^0}}+x_{_{\chi_j^\pm}}\over3}+\varrho_{_{2,1}}(x_{_{\chi_i^0}},x_{_{\chi_j^\pm}})
\nonumber\\
&&\hspace{1.4cm}
+(x_{_{\chi_i^0}}+x_{_{\chi_j^\pm}})\ln x_{_{\rm R}}\Big)+{11\over36x_{_{\rm w}}}
\Big]({\cal O}_{_2}^-+{\cal O}_{_3}^-)
\nonumber\\
&&\hspace{1.4cm}
+\Big(\zeta^{L*}_{_{ij}}\zeta^R_{_{ij}}+\zeta^{R*}_{_{ij}}\zeta^L_{_{ij}}\Big)
(x_{_{\chi_i^0}}x_{_{\chi_j^\pm}})^{1/2}\Big[{5\over12x_{_{\rm w}}^2}\Big(
{1\over\varepsilon}+{5\over6}-\varrho_{_{1,1}}(x_{_{\chi_i^0}},x_{_{\chi_j^\pm}})
\nonumber\\
&&\hspace{1.4cm}
-\ln x_{_{\rm R}}\Big)\Big]
({\cal O}_{_2}^-+{\cal O}_{_3}^-)\Bigg\}+\cdots\;.
\label{w-counter}
\end{eqnarray}

Finally, we get the renormalized effective Lagrangian from the $WW$ sector:
\begin{eqnarray}
&&{\cal L}_{_{WW}}=
-{e^4\over48(4\pi)^2s_{_{\rm w}}^4Q_{_f}\Lambda^2}
\Big(\zeta^{L*}_{_{ij}}\zeta^L_{_{ij}}
+\zeta^{R*}_{_{ij}}\zeta^R_{_{ij}}\Big)
\Big[T_{10}(x_{_{\rm w}},x_{_{\chi_i^0}},x_{_{\chi_j^\pm}})
\nonumber\\
&&\hspace{1.4cm}
+{10\over x_{_{\rm w}}^2}(x_{_{\chi_i^0}}+x_{_{\chi_j^\pm}})\ln x_{_{\rm R}}
\Big]({\cal O}_{_2}^-+{\cal O}_{_3}^-)
\nonumber\\
&&\hspace{1.4cm}
-{e^4\over16(4\pi)^2s_{_{\rm w}}^4Q_{_f}\Lambda^2}
\Big(\zeta^{L*}_{_{ij}}\zeta^L_{_{ij}}
-\zeta^{R*}_{_{ij}}\zeta^R_{_{ij}}\Big)
T_{11}(x_{_{\rm w}},x_{_{\chi_i^0}},x_{_{\chi_j^\pm}})
({\cal O}_{_2}^-+{\cal O}_{_3}^-)
\nonumber\\
&&\hspace{1.4cm}
-{e^4(x_{_{\chi_i^0}}x_{_{\chi_j^\pm}})^{1/2}\over48(4\pi)^2s_{_{\rm w}}^4Q_{_f}\Lambda^2}
\Big(\zeta^{L*}_{_{ij}}
\zeta^R_{_{ij}}+\zeta^{R*}_{_{ij}}\zeta^L_{_{ij}}\Big)
\Big[T_{12}(x_{_{\rm w}},x_{_{\chi_i^0}},x_{_{\chi_j^\pm}})
-{20\over x_{_{\rm w}}^2}\ln x_{_{\rm R}}\Big]({\cal O}_{_2}^-+{\cal O}_{_3}^-)
\nonumber\\
&&\hspace{1.4cm}
-{e^4(x_{_{\chi_i^0}}x_{_{\chi_j^\pm}})^{1/2}\over16(4\pi)^2s_{_{\rm w}}^4Q_{_f}\Lambda^2}
\Big(\zeta^{R*}_{_{ij}}
\zeta^L_{_{ij}}-\zeta^{L*}_{_{ij}}\zeta^R_{_{ij}}\Big)
T_{13}(x_{_{\rm w}},x_{_{\chi_i^0}},x_{_{\chi_j^\pm}})
({\cal O}_{_2}^--{\cal O}_{_3}^-)\;.
\label{ww}
\end{eqnarray}
Correspondingly, the resulted lepton MDMs and EDMs are respectively formulated
as
\begin{eqnarray}
&&a_{l}^{WW}=
-{e^4m_{_l}^2\over12(4\pi)^4s_{_{\rm w}}^4\Lambda^2}
\Big(|\zeta^L_{_{ij}}|^2+|\zeta^R_{_{ij}}|^2\Big)
\Big[T_{10}(x_{_{\rm w}},x_{_{\chi_i^0}},x_{_{\chi_j^\pm}})
\nonumber\\
&&\hspace{1.4cm}
+{10\over x_{_{\rm w}}^2}(x_{_{\chi_i^0}}+x_{_{\chi_j^\pm}})\ln x_{_{\rm R}}
-{32\over x_{_{\rm w}}}\ln x_{_{\rm R}}\Big]
\nonumber\\
&&\hspace{1.4cm}
-{e^4m_{_l}^2\over4(4\pi)^4s_{_{\rm w}}^4\Lambda^2}
\Big(|\zeta^L_{_{ij}}|^2-|\zeta^R_{_{ij}}|^2\Big)
T_{11}(x_{_{\rm w}},x_{_{\chi_i^0}},x_{_{\chi_j^\pm}})
\nonumber\\
&&\hspace{1.4cm}
-{e^4m_{_l}^2(x_{_{\chi_i^0}}x_{_{\chi_j^\pm}})^{1/2}\over6(4\pi)^4s_{_{\rm w}}^4\Lambda^2}
\Re(\zeta^{R*}_{_{ij}}\zeta^L_{_{ij}})
\Big[T_{12}(x_{_{\rm w}},x_{_{\chi_i^0}},x_{_{\chi_j^\pm}})
-{20\over x_{_{\rm w}}^2}\ln x_{_{\rm R}}\Big]\;,
\nonumber\\
&&d_{l}^{WW}=
-{e^5m_{_l}(x_{_{\chi_i^0}}x_{_{\chi_j^\pm}})^{1/2}\over4(4\pi)^4s_{_{\rm w}}^4\Lambda^2}
\Im(\zeta^{R*}_{_{ij}}\zeta^L_{_{ij}})
T_{13}(x_{_{\rm w}},x_{_{\chi_i^0}},x_{_{\chi_j^\pm}})\;.
\label{MEDM-ww}
\end{eqnarray}
In a similar way, the corrections from this sector to the MDM of lepton
depend on a linear combination of the real effective couplings
$|\zeta^L_{_{ij}}|^2\pm|\zeta^R_{_{ij}}|^2$ and real parts of the effective
couplings $\zeta^L_{_{ij}}\zeta^{R*}_{_{ij}}$, and the corrections from this sector
to the EDM of lepton are proportional to imaginary parts of the effective
couplings $\zeta^L_{_{ij}}\zeta^{R*}_{_{ij}}$.

\section{Numerical results and discussion\label{sec3}}
\indent\indent
With the theoretical formulae derived in
previous section, we numerically analyze the dependence of the muon MDM
and EDM on the supersymmetric parameters in the CP-violating scenario
here. In particular, we will present the dependence of the muon MDM
and EDM on the supersymmetric $CP$ phases in some detail. In order to make the
theoretical predictions on the electron and neutron EDMs satisfying
the present experimental constraints, we adopt the cancelation mechanism
among the different contributions to the fermion EDMs \cite{cp3}.Within
three standard error deviations, the present experimental data can
tolerate new physics correction to the muon MDM as
$2.6\times10^{-10}<\Delta a_\mu <58.4\times10^{-10}$. Since the
neutralinos $\chi_{i}^0\;(i=1,\;2,\;3,\;4)$
and charginos $\chi_{i}^\pm\;(i=1,\;2)$ appear
as the internal intermediate particles in the two-loop diagrams
which are investigated in this work, the corrections of these
diagrams will be suppressed strongly when the masses of neutralinos and
charginos are much higher than the electroweak scale\cite{heinemeyer2}.
To investigate if those diagrams can result in concrete corrections to
the muon MDM and EDM, we choose a suitable supersymmetric parameter region where
the masses of neutralinos and charginos are lying in the
range $M_{_{\chi}}<600\;{\rm GeV}$.

The MSSM Lagrangian contains several  sources for  CP violating
phases: the phases of the $\mu$-parameter in the superpotential
and the corresponding bilinear
coupling of the soft breaking terms, three phases of the gaugino
masses, and the phases of the scalar fermion Yukawa couplings in
the soft Lagrangian. As we do not consider the spontaneous CP
violation in this work, the CP phase of soft bilinear coupling
vanishes due to the neutral Higgs tadpole conditions.
Additional, the CP violation would cause changes to the neutral-Higgs-quark
coupling, the neutral Higgs-gauge-boson coupling and the self-coupling of
Higgs boson. A direct result of above facts is that no absolute limits can
be set for the Higgs bosons masses from the present combined LEP data \cite{Bechtle}.
For security, we take the lower bound on the mass of the lightest Higgs boson
as $m_{_{h_1}}\ge60 {\rm GeV}$ \cite{Pilaftsis} in the numerical analysis.
In order to obtain the mixing matrix of neutral Higgs in CP violating MSSM, we include
the subroutine {\it fillhiggs.f} from the Package {\textsf CPsuperH} \cite{JS_Lee}
in our numerical code. Furthermore, we take the pole mass of top quark
$m_t(pole)=175\;{\rm GeV}$, the pole mass of charged Higgs
$m_{_{H^\pm}}(pole)=300\;{\rm GeV}$, the running masses $m_b(m_t)=3\;{\rm GeV},\;
m_\tau(m_t)=1.77{\rm GeV}$, the mass parameters of scalar fermions in
soft terms as $m_{_{{\tilde U}_3}}=m_{_{{\tilde D}_3}}=m_{_{{\tilde E}_3}}
=m_{_{{\tilde Q}_3}}=m_{_{{\tilde L}_3}}=500\;{\rm GeV}$, the Yukawa couplings
of scalar fermions as $|A_t|=|A_b|=|A_\tau|=1\;{\rm TeV}$
and $\phi_{A_t}=\phi_{A_b}=\phi_{A_\tau}=\pi/2$. Fixing above parameters and
assuming $\tan\beta\ge3$, we find that the mass of the lightest neutral Higgs
is well above $115\;{\rm GeV}$ by scanning the parameter space of CP violating MSSM.
In other words, one no longer worries about the constraint from Higgs sector
with the above assumptions on the parameter space of CP violating MSSM.
With no loss of generality, we also take the
supersymmetric parameters $|m_1|=|m_2|=500\;{\rm GeV}$ and $m_{_{{\tilde E}_2}}
=m_{_{{\tilde L}_2}}=A_\mu/2=500\;{\rm GeV}$ in this work.

\begin{figure}[t]
\setlength{\unitlength}{1mm}
\begin{center}
\begin{picture}(0,80)(0,0)
\put(-50,-15){\includegraphics{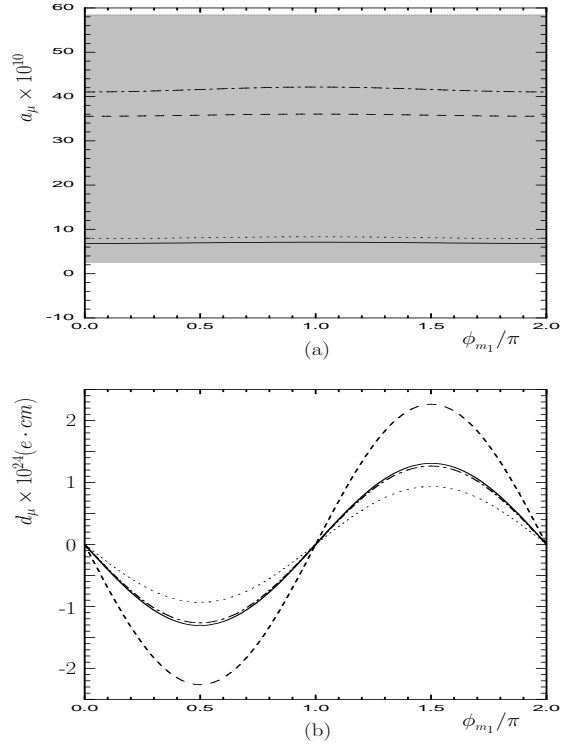}}
\end{picture}
\caption[]{The supersymmetric corrections to the muon MDM $a_\mu$ and
EDM $d_\mu$ vary with the CP violating phase $\phi_{_{m_1}}$
when $|\mu_{_H}|=200\;{\rm GeV},\;\phi_{_{m_2}}=\phi_{_{\mu_{_H}}}=0$ and $\tan\beta=10,\;50$,
where the solid lines stand for the one-loop corrections with $\tan\beta=10$,
the dot lines stand for the results including two-loop supersymmetric
corrections with $\tan=10$; the dash lines stand for the one-loop corrections
with $\tan\beta=50$, the dash-dot lines stand for the results including
two-loop supersymmetric corrections with $\tan=50$. The gray band
in diagram (a) is the region allowed by the $g-2$ experimental data within 3 standard errors.}
\label{fig4}
\end{center}
\end{figure}

Taking $|\mu_{_H}|=200\;{\rm GeV},\;\phi_{_{m_2}}=\phi_{_{\mu_{_H}}}=0$ and $\tan\beta=10,\;50$,
we plot the muon MDM $a_\mu$ and EDM $d_\mu$ versus the CP phase $\phi_{_{m_1}}$ in Fig.\ref{fig4}.
As $\tan\beta=10$, the one-loop supersymmetric correction to the muon
MDM (solid-line in Fig.\ref{fig4}(a)) reaches $7\times10^{-10}$ and can account for
the deviation between the SM prediction and experimental data.
Comparing with one-loop supersymmetric contribution, two-loop contribution depends on
the supersymmetric parameters in a different manner. Including the two-loop corrections,
the supersymmetric contribution to the muon MDM $a_\mu$ is modified about $10\%$.
Since the gaugino mass $m_1$ affects the theoretical prediction only through the mixing
matrix of neutralinos, the muon MDM $a_\mu$ varies with the CP phase $\phi_{_{m_1}}$
(solid line for one-loop result and dot line for the result
including two-loop corrections in Fig.\ref{fig4}(a)) very mildly.
Meanwhile the supersymmetric contribution to the muon EDM including
two-loop corrections at the largest CP violation $\phi_{_{m_1}}=\pi/2$ is
still below $10^{-24} e\cdot cm$ (dot line Fig.\ref{fig4}(b)), and
it is very difficult to observe the muon EDM of this level in next
generation experiments with precision $10^{-24}\;e\cdot cm$ \cite{nexp}.
As $\tan\beta=50$, one-loop supersymmetric correction to the muon MDM
$a_\mu$ exceeds $35\times10^{-10}$ (dash line in Fig.\ref{fig4}(a)),
and can ameliorate easily the discrepancy between the SM prediction
and experiment. Because the dominant two-loop supersymmetric corrections
originating from the $\gamma h_k,\;W^\pm H^\mp$ sectors are enhanced by large $\tan\beta$,
the relative modification from two-loop supersymmetric corrections to one-loop result
is $15\%$ roughly (dash-dot line in Fig.\ref{fig4}(a)). As for the muon EDM $d_\mu$,
one-loop supersymmetric result together with two-loop supersymmetric corrections
are all enhanced by large $\tan\beta$. The contribution including
two-loop supersymmetric corrections is well above $10^{-24} e\cdot cm$
at the largest CP violation $\phi_{_{m_1}}=\pi/2$,
and it is hopeful to detect the muon EDM $d_\mu$ of this level in the near future.

\begin{figure}[t]
\setlength{\unitlength}{1mm}
\begin{center}
\begin{picture}(0,80)(0,0)
\put(-50,-15){\includegraphics{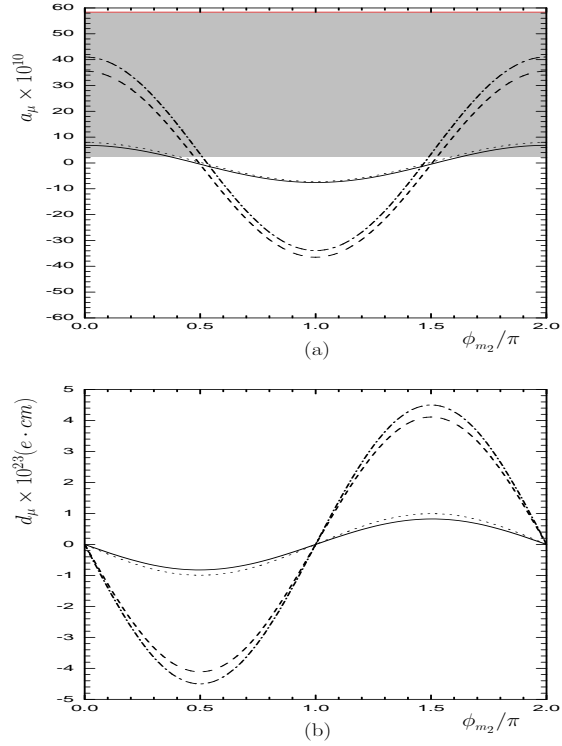}}
\end{picture}
\caption[]{The supersymmetric corrections to the muon MDM $a_\mu$ and
EDM $d_\mu$ vary with the CP violating phase $\phi_{_{m_2}}$
when $|\mu_{_H}|=200\;{\rm GeV},\;\phi_{_{m_1}}=\phi_{_{\mu_{_H}}}=0$ and $\tan\beta=10,\;50$,
where the solid lines stand for the one-loop corrections with $\tan\beta=10$,
the dot lines stand for the results including two-loop supersymmetric
corrections with $\tan=10$; the dash lines stand for the one-loop corrections
with $\tan\beta=50$, the dash-dot lines stand for the results including
two-loop supersymmetric corrections with $\tan=50$. The gray band
in diagram (a) is the region allowed by the $g-2$ experimental data within 3 standard errors.}
\label{fig5}
\end{center}
\end{figure}

Taking $|\mu_{_H}|=200\;{\rm GeV},\;\phi_{_{m_1}}=\phi_{_{\mu_{_H}}}=0$ and $\tan\beta=10,\;50$,
we plot the muon MDM $a_\mu$ and EDM $d_\mu$ versus the CP phase $\phi_{_{m_2}}$ in Fig.\ref{fig5}.
As $\tan\beta=10$, the one-loop supersymmetric correction to the muon
MDM (solid-line in Fig.\ref{fig5}(a)) always lies in the range $|a_\mu|<8\times10^{-10}$
varying with the CP phase  $\phi_{_{m_2}}$. The relative modification from
the two-loop supersymmetric corrections
to the one-loop prediction is below $5\%$ when $\tan\beta=10$.
Since the gaugino mass $m_2$ affects the theoretical prediction through the mixing
matrices of neutralinos and charginos simultaneously, the muon MDM $a_\mu$ depends
on the CP phase $\phi_{_{m_2}}$ (solid line for one-loop result and dot line for the result
including two-loop corrections in Fig.\ref{fig5}(a)) strongly.
Meanwhile the supersymmetric contribution to the muon EDM including
two-loop corrections at the largest CP violation $\phi_{_{m_2}}=\pi/2$ is about $10^{-23} e\cdot cm$
(dot line Fig.\ref{fig5}(b)) which can be observed in next
generation experiments with precision $10^{-24}\;e\cdot cm$ \cite{nexp}.
When $\tan\beta=50$, one-loop supersymmetric correction to the muon MDM
$a_\mu$ is enhanced drastically. Because the dominant two-loop supersymmetric corrections
originating from the $\gamma h_k,\;W^\pm H^\mp$ sectors are also enhanced by large $\tan\beta$,
the relative modification from two-loop supersymmetric corrections to one-loop result
is $15\%$ roughly (dash-dot line in Fig.\ref{fig5}(a)). As for the muon EDM $d_\mu$,
one-loop supersymmetric result together with two-loop supersymmetric corrections
are all enhanced by large $\tan\beta$. The contribution including
two-loop supersymmetric corrections  at the largest CP violation $\phi_{_{m_2}}=\pi/2$
is about $4\times10^{-23} e\cdot cm$ which can be detected easily in next generation
experiments.

\begin{figure}[t]
\setlength{\unitlength}{1mm}
\begin{center}
\begin{picture}(0,80)(0,0)
\put(-50,-15){\includegraphics{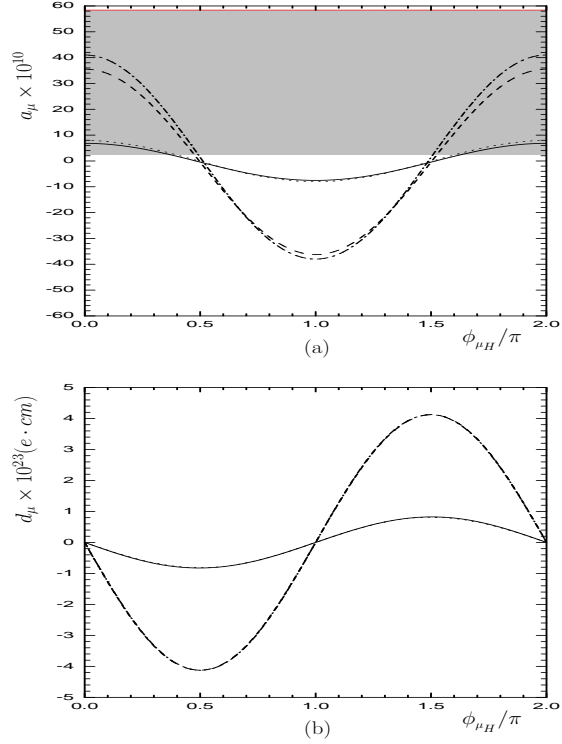}}
\end{picture}
\caption[]{The supersymmetric corrections to the muon MDM $a_\mu$ and
EDM $d_\mu$ vary with the CP violating phase $\phi_{_{\mu_{_H}}}$
when $|\mu_{_H}|=200\;{\rm GeV},\;\phi_{_{m_1}}=\phi_{_{m_2}}=0$ and $\tan\beta=10,\;50$,
where the solid lines stand for the one-loop corrections with $\tan\beta=10$,
the dot lines stand for the results including two-loop supersymmetric
corrections with $\tan=10$; the dash lines stand for the one-loop corrections
with $\tan\beta=50$, the dash-dot lines stand for the results including
two-loop supersymmetric corrections with $\tan=50$. The gray band
in diagram (a) is the region allowed by the $g-2$ experimental data
within 3 standard errors.}
\label{fig6}
\end{center}
\end{figure}

Taking $|\mu_{_H}|=200\;{\rm GeV},\;\phi_{_{m_1}}=\phi_{_{m_2}}=0$ and $\tan\beta=10,\;50$,
we plot the muon MDM $a_\mu$ and EDM $d_\mu$ versus the CP phase $\phi_{_{\mu_{_H}}}$ in Fig.\ref{fig6}.
As $\tan\beta=10$, the one-loop supersymmetric correction to the muon
MDM (solid-line in Fig.\ref{fig6}(a)) always lies in the range $|a_\mu|<8\times10^{-10}$
varying with the CP phase  $\phi_{_{\mu_{_H}}}$. The relative modification from
the two-loop supersymmetric corrections
to the one-loop prediction is below $5\%$ when $\tan\beta=10$.
Since the $\mu$ parameter $\mu_{_H}$ affects the theoretical prediction through the mixing
matrices of neutralinos and charginos simultaneously, the muon MDM $a_\mu$ varies
with the CP phase $\phi_{_{\mu_{_H}}}$ (solid line for one-loop result and dot line for the result
including two-loop corrections in Fig.\ref{fig6}(a)) drastically.
Meanwhile the supersymmetric contribution to the muon EDM including
two-loop corrections at the largest CP violation $\phi_{_{\mu_{_H}}}=\pi/2$
is below $10^{-23} e\cdot cm$ (dot line Fig.\ref{fig6}(b)).
Because the dominant two-loop supersymmetric corrections
originating from the $\gamma h_k,\;W^\pm H^\mp$ sectors are enhanced by large $\tan\beta$,
the relative modification from two-loop supersymmetric corrections to one-loop result
is $15\%$ roughly (dash-dot line in Fig.\ref{fig6}(a)) at CP conservation when $\tan\beta=50$.
One-loop supersymmetric correction to the muon EDM $d_\mu$ is enhanced by large $\tan\beta$.
Comparing with one-loop contribution, two-loop corrections are negligible. The contribution including
two-loop supersymmetric corrections is about $4\times10^{-23} e\cdot cm$,
which can be detected in the near future \cite{nexp}.

\begin{figure}[t]
\setlength{\unitlength}{1mm}
\begin{center}
\begin{picture}(0,80)(0,0)
\put(-100,-25){\includegraphics{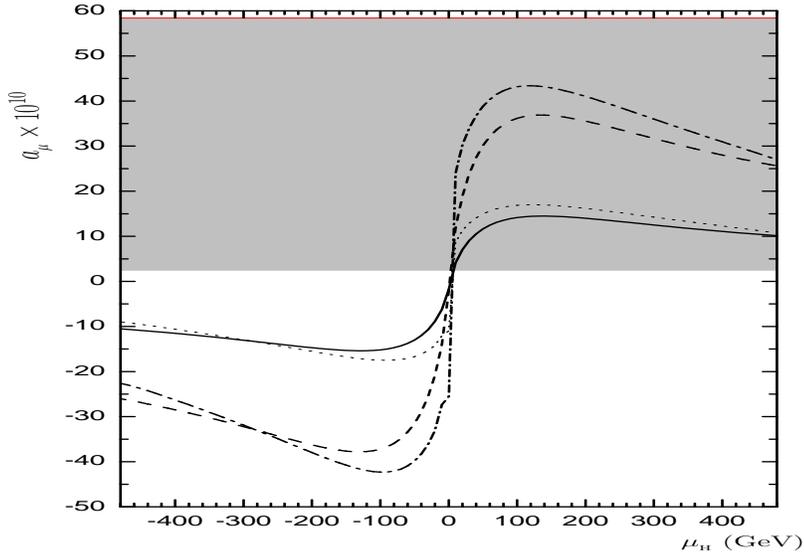}}
\end{picture}
\caption[]{The supersymmetric corrections to the muon MDM $a_\mu$
vary with the $\mu$-parameter $\mu_{_H}$
when $\phi_{_{m_1}}=\phi_{_{m_2}}=\phi_{_{\mu_{_H}}}=0$ and $\tan\beta=20,\;50$,
where the solid lines stand for the one-loop corrections with $\tan\beta=20$,
the dot lines stand for the results including two-loop supersymmetric
corrections with $\tan=20$; the dash lines stand for the one-loop corrections
with $\tan\beta=50$, the dash-dot lines stand for the results including
two-loop supersymmetric corrections with $\tan=50$. The gray band
is the region allowed by the $g-2$ experimental data within 3 standard errors.}
\label{fig7}
\end{center}
\end{figure}

Taking $\tan\beta=20,\;50$ and $\phi_{_{m_1}}=\phi_{_{m_2}}=\phi_{_{\mu_{_H}}}=0$,
we plot the  muon MDM $a_\mu$ versus the $\mu$-parameter $\mu_{_H}$ in Fig.\ref{fig7}.
The gray band is the region allowed by present experimental data within 3 standard errors.
Because the supersymmetric corrections to the muon MDM $a_\mu$ are negative for $\mu_{_H}\le0$,
the corresponding parameter space is already ruled out by the present $g-2$
experimental data. Comparing with the one-loop supersymmetric results
(solid line for $\tan\beta=20$ and dash line for $\tan\beta=50$
respectively), the contributions including two-loop supersymmetric corrections
are enhanced about $15\%$ when $\mu_{_H}=150\;{\rm GeV}$. Along with
the increasing of $\mu_{_H}$, the two-loop corrections become more and more
trivial.

\section{Conclusions\label{sec4}}
\indent\indent
In this work, we analyzed the two-loop
supersymmetric corrections to the muon MDM and EDM
by the effective Lagrangian method in the CP violating MSSM.
In the concrete calculation, we keep all dimension 6 operators. The ultraviolet divergence
caused by the divergent sub-diagrams is removed in the on-shell renormalization
schemes. After applying the equations of motion to the external leptons, we derive the
muon MDM and EDM. Numerically, we analyze the dependence of the muon
MDM $a_\mu$ and EDM $d_\mu$ on supersymmetric CP violating phases.
As discussed above, $a_\mu$ is decided by real parts of the effective couplings,
and $d_\mu$ is decided by imaginary parts of the effective couplings
after the heavy freedoms are integrated out.
Adopting our assumptions on parameter space of the MSSM and
choosing $\tan\beta=50$, we find that the correction from those two-loop diagrams to
$a_\mu$ is $4\times10^{-10}$ roughly for the case of CP conservation, which lies in the order
of present experimental precision in magnitude. In other words,
the present experimental data put a very restrictive bound on the
real parts of those effective couplings. Additional, the contribution to $d_\mu$
from this sector is sizable enough to be experimentally detected with
the experimental precision of near future.

\begin{acknowledgments}
\indent\indent
The work has been supported by the National Natural Science Foundation of China (NNSFC)
with Grant No. 10675027.
\end{acknowledgments}
\vspace{1.6cm}
\appendix

\section{The functions\label{ap1}}
\indent\indent
We list the tedious expressions of the functions adopted in the text
\begin{eqnarray}
&&\varrho_{_{i,j}}(x,y)={x^i\ln^jx-y^i\ln^jy\over x-y}\;,
\nonumber\\
&&\Omega_{_n}(x,y;u,v)={x^n\Phi(x,u,v)-y^n\Phi(y,u,v)\over x-y}\;,
\nonumber\\
&&T_1(x_1,x_2,x_3)={1\over x_1}\Bigg\{-4(2+\ln x_2)(\ln x_1-1)
-{\partial\over\partial x_3}\Big[\Big(1+2{x_2-x_3\over x_1}\Big)\Phi\Big]
(x_1,x_2,x_3)
\nonumber\\
&&\hspace{2.8cm}
+{\partial\over\partial x_3}\Big[\Big(1+2{x_2-x_3\over x_1}\Big)
\varphi_0+2(x_2-x_3)\varphi_1\Big](x_2,x_3)\Bigg\}\;,
\nonumber\\
&&T_2(x_1,x_2,x_3)
={1\over x_1}\Bigg[{\partial \Phi\over\partial x_3}
(x_1,x_2,x_3)-{\partial\varphi_0\over\partial x_3}(x_2,x_3)\Bigg]\;,
\nonumber\\
&&T_3(x_1,x_2,x_3)
=-{2\over x_1}(2+\ln x_3)+{2\over x_1}{\partial^2\over\partial x_3^2}
\Big(x_3\Phi\Big)(x_1,x_2,x_3)
\nonumber\\
&&\hspace{2.8cm}
-{2\over x_1}{\partial^2\over\partial x_3^2}
\Big(x_3\varphi_0\Big)(x_2,x_3)
-{4\over x_1}{\partial\Phi\over\partial x_3}
(x_1,x_2,x_3)
\nonumber\\
&&\hspace{2.8cm}
+{4\over x_1}{\partial\varphi_0\over\partial x_3}
(x_2,x_3)+{\partial^2\over\partial x_1\partial x_3}
\Big({x_2-x_3\over x_1}\varphi_0\Big)(x_2,x_3)
\nonumber\\
&&\hspace{2.8cm}
+{\partial^2\over\partial x_1\partial x_3}
\Big[\Big(1-{x_2-x_3\over x_1}\Big)\Phi\Big](x_1,x_2,x_3)\;,
\nonumber\\
&&T_4(x_1,x_2,x_3)={2\over x_1}\ln x_3-{2\over x_1^2}\Big(x_2-x_2\ln x_2
-x_3+x_3\ln x_3\Big)
\nonumber\\
&&\hspace{2.8cm}
-{\partial^3\over\partial x_1\partial x_3^2}\Big[{x_2x_3-x_3^2\over x_1}
\Big(\Phi(x_1,x_2,x_3)-\varphi_0(x_2,x_3)\Big)\Big]
\nonumber\\
&&\hspace{2.8cm}
+{1\over2}{\partial^3\over\partial x_1^2\partial x_3}
\Big[(x_2-3x_3-x_1)\Phi(x_1,x_2,x_3)\Big]
\nonumber\\
&&\hspace{2.8cm}
-{1\over2}{\partial^2\over\partial x_1\partial x_3}\Big[\Phi(x_1,x_2,x_3)
-{5\over x_1}(x_2-x_3)\Big(\Phi(x_1,x_2,x_3)
\nonumber\\
&&\hspace{2.8cm}
-\varphi_0(x_2,x_3)\Big)\Big]-{\partial^2\over\partial x_1^2}
\Big[{x_2-x_3\over x_1}\Big(\Phi(x_1,x_2,x_3)-\varphi_0(x_2,x_3)\Big)
\nonumber\\
&&\hspace{2.8cm}
+2\Phi(x_1,x_2,x_3)\Big]\;,
\nonumber\\
&&T_5(x_1,x_2,x_3)
={5\over12x_1}+\Big({5\over12x_1^2}+{\ln x_1\over3x_1^2}+{\ln x_{_{\rm R}}\over x_1^2}\Big)(x_2+x_3)
\nonumber\\
&&\hspace{2.8cm}
+\Big({7\over6x_1^2}+{2\over3x_1^2}\ln x_1\Big)(x_2\ln x_2+x_3\ln x_3)
\nonumber\\
&&\hspace{2.8cm}
+\Big({2\over3x_1^3}-{4\over3x_1^3}\ln x_1\Big)(x_2-x_3)^2
(1+\varrho_{_{1,1}}(x_2,x_3))
\nonumber\\
&&\hspace{2.8cm}
+{23\over6x_1^2}(x_2+x_3)\Big(1+\varrho_{_{1,1}}(x_2,x_3)\Big)
-{5\varrho_{_{2,1}}(x_2,x_3)\over x_1^2}
\nonumber\\
&&\hspace{2.8cm}
-{1\over3x_1^2}\Big(1-{2(x_2+x_3)\over x_1}\Big)\Big(\Phi(x_1,x_2,x_3)
-\varphi_0(x_2,x_3)\Big)
\nonumber\\
&&\hspace{2.8cm}
+{1\over3x_1}\Big({x_2+x_3\over x_1}-{2(x_2-x_3)^2\over x_1^2}\Big)\varphi_1(x_2,x_3)
\nonumber\\
&&\hspace{2.8cm}
+{1\over3x_1}\Big(1-{3(x_2+x_3)\over x_1}+{2(x_2-x_3)^2\over x_1^2}\Big)
{\partial\Phi\over\partial x_1}(x_1,x_2,x_3)
\nonumber\\
&&\hspace{2.8cm}
-{1\over3}\Big(1-{2(x_2+x_3)\over x_1}+{(x_2-x_3)^2\over x_1^2}\Big)
{\partial^2\Phi\over\partial x_1^2}(x_1,x_2,x_3)
\nonumber\\
&&\hspace{2.8cm}
-{(x_2-x_3)^2\over3x_1^2}\varphi_2(x_2,x_3)\;,
\nonumber\\
&&T_{6}(x_1,x_2,x_3)
=-{1\over x_1^2}\Big(\varphi_0-(x_2-x_3)
{\partial\varphi_0\over\partial x_3}\Big)(x_2,x_3)
+\Big[2x_3{\partial^3\Phi\over\partial x_1\partial x_3^2}
+{\partial^2\Phi\over\partial x_1^2}
\nonumber\\
&&\hspace{3.0cm}
+(x_1-x_2+x_3){\partial^3\Phi\over\partial x_1^2\partial x_3}
+{\Phi\over x_1^2}-{x_2-x_3\over x_1^2}{\partial\Phi\over\partial x_3}
-{1\over x_1}{\partial\Phi\over\partial x_1}
\nonumber\\
&&\hspace{3.0cm}
+(1+{x_2-x_3\over x_1}){\partial^2\Phi\over\partial x_1\partial x_3}\Big]
(x_1,x_2,x_3)\;,
\nonumber\\
&&T_{7}(x_1,x_2,x_3)
=-2{\partial^3\Phi\over\partial x_1^2\partial x_3}(x_1,x_2,x_3)
+{2\over x_1x_3}-{2\over x_1^2}\Big(\ln x_2-\ln x_3\Big)
\nonumber\\
&&\hspace{3.0cm}
+\Big({\partial^3\over\partial x_1^2\partial x_3}
-{\partial^3\over\partial x_1\partial x_3^2}
+{\partial^3\over\partial x_1^2\partial x_2}
+{\partial^3\over\partial x_1\partial x_2\partial x_3}\Big)\Big[\Phi(x_1,x_2,x_3)
\nonumber\\
&&\hspace{3.0cm}
-{x_2-x_3\over x_1}\Big(\Phi(x_1,x_2,x_3)-\varphi_0(x_2,x_3)\Big)\Big]\;,
\nonumber\\
&&T_{8}(x_1,x_2,x_3)
=-4\Big({\partial^3\Phi\over\partial x_1^2\partial x_3}
+{\partial^3\Phi\over\partial x_1^2\partial x_2}\Big)(x_1,x_2,x_3)
+{4\over x_1x_3}+{2\over x_1^2}(2+\ln x_2)
\nonumber\\
&&\hspace{3.0cm}
+\Big(2{\partial^3\over\partial x_1\partial x_3^2}
+{\partial^3\over\partial x_1^2\partial x_2}\Big)
\Big[{x_2-x_3\over x_1}\Big(\Phi(x_1,x_2,x_3)-\varphi_0(x_2,x_3)\Big)
\nonumber\\
&&\hspace{3.0cm}
-\Phi(x_1,x_2,x_3)\Big]\;,
\nonumber\\
&&T_{9}(x_1,x_2,x_3)={2\over x_1}\ln x_3
-{4x_3\over x_1^2}\Big({\partial\Phi\over\partial x_3}(x_1,x_2,x_3)
-{\partial\varphi_0\over\partial x_3}(x_2,x_3)\Big)
\nonumber\\
&&\hspace{3.0cm}
+{\partial^2\over\partial x_1\partial x_3}\Big((x_2-x_3)
{\Phi(x_1,x_2,x_3)-\varphi_0(x_2,x_3)\over x_1}-\Phi(x_1,x_2,x_3)\Big)
\nonumber\\
&&\hspace{3.0cm}
+{4\over x_1}\Big({\partial\Phi\over\partial x_3}
-{\partial\Phi\over\partial x_1}\Big)(x_1,x_2,x_3)
+{4x_3\over x_1}{\partial^2\Phi\over\partial x_1\partial x_3}(x_1,x_2,x_3)\;,
\nonumber\\
&&T_{10}(x_1,x_2,x_3)
={26\over x_1}+{17x_2\over x_1^2}+{29x_3\over x_1^2}
+{10\over x_1^2}\varrho_{_{2,1}}(x_2,x_3)
-{16(x_2-x_3)^2\over x_1^3}
\nonumber\\
&&\hspace{3.0cm}
-{10(x_2+x_3)\over x_1^2}\ln x_1-{6\ln x_3\over x_1}
+\Big[14-{16(x_2-x_3)\over x_1}
\Big]{x_2\ln x_2\over x_1^2}
\nonumber\\
&&\hspace{3.0cm}
+\Big[-4+{16(x_2-x_3)\over x_1}\Big]{x_3\ln x_3\over x_1^2}
+\Big[(x_2-x_3)^2-x_1^2\Big]{\partial^4\Phi\over\partial x_1^4}(x_1,x_2,x_3)
\nonumber\\
&&\hspace{3.0cm}
+\Big[-5x_1+6x_2+{3(x_2-x_3)^2\over x_1}\Big]
{\partial^3\Phi\over\partial x_1^3}(x_1,x_2,x_3)
\nonumber\\
&&\hspace{3.0cm}
+\Big[-{9(x_2-x_3)^2\over x_1^2}+{6x_2\over x_1}
+{3x_3\over x_1}\Big]{\partial^2\Phi\over\partial x_1^2}(x_1,x_2,x_3)
\nonumber\\
&&\hspace{3.0cm}
+\Big[-{12x_2\over x_1^2}-{6x_3\over x_1^2}+{18(x_2-x_3)^2\over x_1^3}\Big]
{\partial\Phi\over\partial x_1}(x_1,x_2,x_3)
\nonumber\\
&&\hspace{3.0cm}
+\Big[{12x_2\over x_1^3}+{6x_3\over x_1^3}-{18(x_2-x_3)^2\over x_1^4}\Big]
\Big(\Phi(x_1,x_2,x_3)-\varphi_0(x_2,x_3)\Big)
\nonumber\\
&&\hspace{3.0cm}
+{2x_3^2(x_2-x_3)\over x_1^2}\Big[{\partial^3\Phi\over\partial x_3^3}
(x_1,x_2,x_3)-{\partial^3\varphi_0\over\partial x_3^3}(x_2,x_3)\Big]
\nonumber\\
&&\hspace{3.0cm}
+\Big[{3x_\alpha x_\beta\over x_1^2}-{9x_\beta^2\over x_1^2}\Big]
\Big[{\partial^2\Phi\over\partial x_3^2}(x_1,x_2,x_3)
-{\partial^2\varphi_0\over\partial x_3^2}(x_2,x_3)\Big]
\nonumber\\
&&\hspace{3.0cm}
-\Big[{3x_\alpha\over x_1^2}+{9x_\beta\over x_1^2}+{18x_3(x_2-x_3)
\over x_1^3}\Big]\Big[{\partial\Phi\over\partial x_3}(x_1,x_2,x_3)
\nonumber\\
&&\hspace{3.0cm}
-{\partial\varphi_0\over\partial x_3}(x_2,x_3)\Big]-6x_3(x_2-x_3+x_1)
{\partial^4\Phi\over\partial x_1^3\partial x_3}(x_1,x_2,x_3)
\nonumber\\
&&\hspace{3.0cm}
+6x_3(x_2+x_3-x_1){\partial^4\Phi\over\partial x_1^2\partial x_3^2}
(x_1,x_2,x_3)
\nonumber\\
&&\hspace{3.0cm}
-2x_3^2\Big(1+{x_2-x_3\over x_1}\Big)
{\partial^4\Phi\over\partial x_1\partial x_3^3}(x_1,x_2,x_3)
\nonumber\\
&&\hspace{3.0cm}
+\Big[3x_1-3x_2-18x_3-{9x_3(x_2-x_3)\over x_1}\Big]
{\partial^3\Phi\over\partial x_1^2\partial x_3}(x_1,x_2,x_3)
\nonumber\\
&&\hspace{3.0cm}
+\Big[-21x_3-{3x_2x_3\over x_1}+{9x_3^2\over x_1}\Big]
{\partial^3\Phi\over\partial x_1\partial x_3^2}(x_1,x_2,x_3)
\nonumber\\
&&\hspace{3.0cm}
-\Big[6-{12x_2\over x_1}+{6x_3\over x_1}-{18x_3(x_2-x_3)\over x_1^2}\Big]
{\partial^2\Phi\over\partial x_1\partial x_3}(x_1,x_2,x_3)\;,
\nonumber\\
&&T_{11}(x_1,x_2,x_3)={2\ln x_3\over x_1}-{4(x_2-x_3)\over x_1^2}
-{4(x_2\ln x_2-x_3\ln x_3)\over x_1^2}
\nonumber\\
&&\hspace{3.0cm}
-{4(x_2-x_3)\over x_1^3}\Big(\Phi(x_1,x_2,x_3)-\varphi_0(x_2,x_3)\Big)
+{4(x_2-x_3)\over x_1^2}{\partial\Phi\over\partial x_1}(x_1,x_2,x_3)
\nonumber\\
&&\hspace{3.0cm}
-\Big(1+{2(x_2-x_3)\over x_1}\Big){\partial^2\Phi\over\partial x_1^2}
(x_1,x_2,x_3)
-{2x_3\over x_1^2}\Big({\partial\Phi\over\partial x_3}(x_1,x_2,x_3)
\nonumber\\
&&\hspace{3.0cm}
-{\partial\varphi_0\over\partial x_3}(x_2,x_3)\Big)+{x_3(x_2-x_3)\over x_1^2}
\Big({\partial^2\Phi\over\partial x_3^2}(x_1,x_2,x_3)
-{\partial^2\varphi_0\over\partial x_3^2}(x_2,x_3)\Big)
\nonumber\\
&&\hspace{3.0cm}
-2{\partial^2\Phi\over\partial x_1\partial x_3}(x_1,x_2,x_3)
-x_3\Big(1+{x_2-x_3\over x_1}\Big)
{\partial^3\Phi\over\partial x_1\partial x_3^2}(x_1,x_2,x_3)
\nonumber\\
&&\hspace{3.0cm}
+\Big(x_2+x_3-x_1\Big){\partial^3\Phi\over\partial x_1^2\partial x_3}
(x_1,x_2,x_3)\;,
\nonumber\\
&&T_{12}(x_1,x_2,x_3)
=-{52\over x_1^2}+{4\over x_1x_3}+{20\over x_1^2}\ln x_1-{18\ln x_3\over x_1^2}
-{20\over x_1^2}\varrho_{_{1,1}}(x_2,x_3)
\nonumber\\
&&\hspace{3.0cm}
-{12\over x_1^3}\Big(\Phi(x_1,x_2,x_3)-\varphi_0(x_2,x_3)\Big)
+{12\over x_1^2}{\partial\Phi\over\partial x_1}(x_1,x_2,x_3)
\nonumber\\
&&\hspace{3.0cm}
-{6\over x_1}{\partial^2\Phi\over\partial x_1^2}(x_1,x_2,x_3)
-\Big(17{\partial^3\Phi\over\partial x_1^3}
+2x_1{\partial^4\Phi\over\partial x_1^4}\Big)(x_1,x_2,x_3)
\nonumber\\
&&\hspace{3.0cm}
+{6\over x_1^2}\Big(1+{2(x_2-x_3)\over x_1}\Big)
\Big({\partial\Phi\over\partial x_3}(x_1,x_2,x_3)
-{\partial\varphi_0\over\partial x_3}(x_2,x_3)\Big)
\nonumber\\
&&\hspace{3.0cm}
-{3(x_2-2x_3)\over x_1^2}\Big({\partial^2\Phi\over\partial x_3^2}
(x_1,x_2,x_3)-{\partial^2\varphi_0\over\partial x_3^2}(x_2,x_3)\Big)
\nonumber\\
&&\hspace{3.0cm}
-{x_3(x_2-x_3)\over x_1^2}\Big({\partial^3\Phi\over\partial x_3^3}
(x_1,x_2,x_3)-{\partial^3\varphi_0\over\partial x_3^3}(x_2,x_3)\Big)
\nonumber\\
&&\hspace{3.0cm}
-x_3\Big(1-{x_2-x_3\over x_1}\Big){\partial^4\Phi\over\partial x_1\partial x_3^3}
(x_1,x_2,x_3)
\nonumber\\
&&\hspace{3.0cm}
-{6\over x_1}\Big(1+{2(x_2-x_3)\over x_1}\Big)
{\partial^2\Phi\over\partial x_1\partial x_3}(x_1,x_2,x_3)
\nonumber\\
&&\hspace{3.0cm}
-\Big[3\Big(1-{x_2-2x_3\over x_1}\Big){\partial^3\Phi\over\partial x_1\partial x_3^2}
+6\Big(2-{x_2-x_3\over x_1}\Big){\partial^3\Phi\over\partial x_1^2\partial x_3}\Big]
(x_1,x_2,x_3)
\nonumber\\
&&\hspace{3.0cm}
+3(x_2-x_3-x_1){\partial^4\Phi\over\partial x_1^3\partial x_3}
(x_1,x_2,x_3)-6{\partial^4\Phi\over\partial x_1^2\partial x_3^2}(x_1,x_2,x_3)\;,
\nonumber\\
&&T_{13}(x_1,x_2,x_3)
={1\over x_1x_3}+{2\over x_1^2}\Big({\partial\Phi\over\partial x_3}
(x_1,x_2,x_3)-{\partial\varphi_0\over\partial x_3}(x_2,x_3)\Big)
-{2\over x_1}{\partial^2\Phi\over\partial x_1\partial x_3}(x_1,x_2,x_3)
\nonumber\\
&&\hspace{3.0cm}
-{x_2-x_3\over x_1^2}
\Big({\partial^2\Phi\over\partial x_3^2}
(x_1,x_2,x_3)
-{\partial^2\varphi_0\over\partial x_3^2}(x_2,x_3)\Big)
\nonumber\\
&&\hspace{3.0cm}
-\Big(1-{x_2-x_3\over x_1}\Big)
{\partial^3\Phi\over\partial x_1\partial x_3^2}(x_1,x_2,x_3)
-2{\partial^3\Phi\over\partial x_1^2\partial x_3}(x_1,x_2,x_3)\;,
\nonumber\\
&&F_1(x_1,x_2,x_3,x_4)={1\over x_1x_2}{\partial\over\partial x_4}\Big((x_3-x_4)\varphi_0\Big)(x_3,x_4)
\nonumber\\
&&\hspace{3.6cm}
+{1\over x_1-x_2}\Big\{{\partial\over\partial
x_4}\Big[\Big(1+{x_3-x_4\over x_1}\Big)\Phi\Big](x_1,x_3,x_4)
\nonumber\\
&&\hspace{3.6cm}
-{\partial\over\partial x_4}\Big[\Big(1+{x_3-x_4\over x_2}\Big)\Phi\Big](x_2,x_3,x_4)\Big\}\;,
\nonumber\\
&&F_2(x_1,x_2,x_3,x_4)=-{1\over x_1x_2}{\partial\over\partial x_4}\Big((x_3-x_4)\varphi_0\Big)(x_3,x_4)
\nonumber\\
&&\hspace{3.6cm}
+{1\over x_1-x_2}\Big\{{\partial\over\partial
x_4}\Big[\Big(1-{x_3-x_4\over x_1}\Big)\Phi\Big](x_1,x_3,x_4)
\nonumber\\
&&\hspace{3.6cm}
-{\partial\over\partial x_4}\Big[\Big(1-{x_3-x_4\over x_2}\Big)\Phi\Big](x_2,x_3,x_4)\Big\}\;,
\nonumber\\
&&F_3(x_1,x_2,x_3,x_4)=
2(\ln x_4-1)\varrho_{_{0,1}}(x_1,x_2)-{6(x_3-x_4)\over x_1x_2}
-{6(x_3\ln x_3-x_4\ln x_4)\over x_1x_2}
\nonumber\\
&&\hspace{3.6cm}
+{x_1x_2+2(x_1+x_2)(x_3-x_4)\over x_1^2x_2^2}\varphi_0(x_3,x_4)
-{x_3-3x_4\over x_1x_2}{\partial\varphi_0\over\partial x_4}(x_3,x_4)
\nonumber\\
&&\hspace{3.6cm}
-{x_4(x_3-x_4)\over x_1x_2}
{\partial^2\varphi_0\over\partial x_4^2}(x_3,x_4)
-\Big({\partial\over\partial x_4}+x_4{\partial^2\over\partial x_4^2}\Big)
\Omega_{_0}(x_1,x_2;x_3,x_4)
\nonumber\\
&&\hspace{3.6cm}
+\Big(1-(x_3-3x_4){\partial\over\partial x_4}-x_4
(x_3-x_4){\partial^2\over\partial x_4^2}\Big)\Omega_{_{-1}}(x_1,x_2;x_3,x_4)
\nonumber\\
&&\hspace{3.6cm}
-\Big({\partial\over\partial x_1}+{\partial\over\partial x_2}\Big)^2
\Big[\Omega_{_1}(x_1,x_2;x_3,x_4)
+(x_3-x_4)\Omega_{_0}(x_1,x_2;x_3,x_4)\Big]
\nonumber\\
&&\hspace{3.6cm}
-2\Big({\partial\over\partial x_1}+{\partial\over\partial x_2}\Big)
\Big[{\partial\Omega_{_1}\over\partial x_4}(x_1,x_2;x_3,x_4)
-(x_3+x_4){\partial\Omega_{_0}\over\partial x_4}(x_1,x_2;x_3,x_4)\Big]
\nonumber\\
&&\hspace{3.6cm}
-2(x_3-x_4)\Big({\partial\over\partial x_1}+{\partial\over\partial x_2}\Big)
\Omega_{_{-1}}(x_1,x_2;x_3,x_4)\;,
\nonumber\\
&&F_4(x_1,x_2,x_3,x_4)=
2(\ln x_4-1)\varrho_{_{0,1}}(x_1,x_2)-{6(x_3-x_4)\over x_1x_2}
-{6(x_3\ln x_3-x_4\ln x_4)\over x_1x_2}
\nonumber\\
&&\hspace{3.6cm}
-{x_1x_2-2(x_1+x_2)(x_3-x_4)\over x_1^2x_2^2}\varphi_0(x_3,x_4)
+{x_3+x_4\over x_1x_2}{\partial\varphi_0\over\partial x_4}(x_3,x_4)
\nonumber\\
&&\hspace{3.6cm}
-{x_4(x_3-x_4)\over x_1x_2}
{\partial^2\varphi_0\over\partial x_4^2}(x_3,x_4)
+\Big(-{\partial\over\partial x_4}+x_4{\partial^2\over\partial x_4^2}\Big)
\Omega_{_0}(x_1,x_2;x_3,x_4)
\nonumber\\
&&\hspace{3.6cm}
+\Big(-1+(x_3+x_4){\partial\over\partial x_4}-x_4
(x_3-x_4){\partial^2\over\partial x_4^2}\Big)\Omega_{_{-1}}(x_1,x_2;x_3,x_4)
\nonumber\\
&&\hspace{3.6cm}
+\Big({\partial\over\partial x_1}+{\partial\over\partial x_2}\Big)^2
\Big[\Omega_{_1}(x_1,x_2;x_3,x_4)
-(x_3-x_4)\Omega_{_0}(x_1,x_2;x_3,x_4)\Big]
\nonumber\\
&&\hspace{3.6cm}
-2\Big({\partial\over\partial x_1}+{\partial\over\partial x_2}\Big)
\Big[\Omega_{_0}(x_1,x_2;x_3,x_4)
-2x_4{\partial\Omega_{_0}\over\partial x_4}(x_1,x_2;x_3,x_4)\Big]
\nonumber\\
&&\hspace{3.6cm}
-2(x_3-x_4)\Big({\partial\over\partial x_1}+{\partial\over\partial x_2}\Big)
\Omega_{_{-1}}(x_1,x_2;x_3,x_4)\;,
\nonumber\\
&&F_5(x_1,x_2,x_3,x_4)=-2(2+\ln x_4)\varrho_{_{0,1}}(x_1,x_2)
+{1\over x_1x_2}\varphi_0(x_3,x_4)
\nonumber\\
&&\hspace{3.6cm}
-{x_3-x_4\over x_1x_2}{\partial\varphi_0\over\partial x_4}(x_3,x_4)
-{\partial\Omega_{_0}\over\partial x_4}(x_1,x_2;x_3,x_4)
\nonumber\\
&&\hspace{3.6cm}
+\Big(1-(x_3-x_4){\partial\over\partial x_4}\Big)\Omega_{-1}
(x_1,x_2;x_3,x_4)\;,
\nonumber\\
&&F_6(x_1,x_2,x_3,x_4)=2(2+\ln x_4)\varrho_{_{0,1}}(x_1,x_2)
-{1\over x_1x_2}\varphi_0(x_3,x_4)
\nonumber\\
&&\hspace{3.6cm}
+{x_3-x_4\over x_1x_2}{\partial\varphi_0\over\partial x_4}(x_3,x_4)
-{\partial\Omega_{_0}\over\partial x_4}(x_1,x_2;x_3,x_4)
\nonumber\\
&&\hspace{3.6cm}
-\Big(1-(x_3-x_4){\partial\over\partial x_4}\Big)\Omega_{-1}
(x_1,x_2;x_3,x_4)\;.
\label{funs}
\end{eqnarray}

The concrete expression of $\Phi(x,y,z)$ can be found in \cite{Feng2,2vac}. In the limit
$z\ll x,y$, we can expand $\Phi(x,y,z)$ according $z$ as
\begin{eqnarray}
&&\Phi(x,y,z)=\varphi_0(x,y)+z\varphi_1(x,y)+{z^2\over2!}\varphi_2(x,y)
+{z^3\over3!}\varphi_3(x,y)+{z^4\over4!}\varphi_4(x,y)
\nonumber\\
&&\hspace{2.2cm}
+2z\Big(\ln z-1\Big)\Big(1+\varrho_{_{1,1}}(x,y)\Big)
\nonumber\\
&&\hspace{2.2cm}
-2z^2\Big({\ln z\over2!}-{3\over4}\Big)
\Big({x+y\over(x-y)^2}+{2xy\over(x-y)^3}\ln{y\over x}\Big)
\nonumber\\
&&\hspace{2.2cm}
-{2z^3\over(x-y)^2}\Big({\ln z\over3!}-{11\over36}\Big)\Big(1+{12xy\over(x-y)^2}
+{6xy(x+y)\over(x-y)^3}\ln{y\over x}\Big)
\nonumber\\
&&\hspace{2.2cm}
-2z^4\Big({\ln z\over4!}-{25\over288}\Big)\Big(
{2x^3+58x^2y+58xy^2+2y^3\over(x-y)^6}
\nonumber\\
&&\hspace{2.2cm}
+{24xy(x^2+3xy+y^2)\over(x-y)^7}\ln{y\over x}\Big)+\cdots
\label{phi-expand}
\end{eqnarray}
with
\begin{eqnarray}
&&\varphi_0(x,y)=\left\{\begin{array}{ll}(x+y)\ln x\ln y+(x-y)\Theta(x,y)
\;,&x>y\;;\\
2x\ln^2x\;,&x=y\;;\\
(x+y)\ln x\ln y+(y-x)\Theta(y,x)
\;,&x<y\;.\end{array}\right.
\label{varphi0}
\end{eqnarray}

\begin{eqnarray}
&&\varphi_1(x,y)=\left\{\begin{array}{ll}-\ln x\ln y-{x+y\over x-y}\Theta(x,y)
\;,&x>y\;;\\
4-2\ln x-\ln^2x\;,&x=y\;;\\
-\ln x\ln y-{x+y\over y-x}\Theta(y,x)
\;,&x<y\;.\end{array}\right.
\label{varphi1}
\end{eqnarray}

\begin{eqnarray}
&&\varphi_2(x,y)=\left\{\begin{array}{ll}
{(2x^2+6xy)\ln x-(6xy+2y^2)\ln y\over(x-y)^3}-{4xy\over(x-y)^3}\Theta(x,y)
\;,&x>y\;;\\
-{5\over9x}+{2\over3x}\ln x\;,&x=y\;;\\
{(2x^2+6xy)\ln x-(6xy+2y^2)\ln y\over(x-y)^3}-{4xy\over(y-x)^3}\Theta(y,x)
\;,&x<y\;.\end{array}\right.
\label{varphi2}
\end{eqnarray}

\begin{eqnarray}
&&\varphi_3(x,y)=\left\{\begin{array}{ll}-{12xy(x+y)\over(x-y)^5}\Theta(x,y)
-{2(x^2+xy+y^2)\over(x-y)^4} & \\
+{2(x^3+14x^2y+11xy^2)\ln x-2(y^3+14xy^2+11x^2y)\ln y\over(x-y)^5}
\;,&x>y\;;\\
-{53\over150x^2}+{1\over5x^2}\ln x\;,&x=y\;;\\
-{12xy(x+y)\over(y-x)^5}\Theta(y,x)-{2(x^2+xy+y^2)\over(x-y)^4} & \\
+{2(x^3+14x^2y+11xy^2)\ln x-2(y^3+14xy^2+11x^2y)\ln y\over(x-y)^5}
\;,&x<y\;.
\end{array}\right.
\label{varphi3}
\end{eqnarray}

\begin{eqnarray}
&&\varphi_4(x,y)=\left\{\begin{array}{ll}
-{48xy(x^2+3xy+y^2)\over(x-y)^7}\Theta(x,y)-{2(3x^3+61x^2y+61xy^2+3y^3)\over(x-y)^6} & \\
+{4(x^4+3x^3y-45x^2y^2-25xy^3)\ln x-4(y^4+3y^3x-45x^2y^2-25yx^3)\ln y\over(x-y)^7}
\;,&x>y\;;\\
-{598\over2205x^3}+{1\over210x^3}\ln x\;,&x=y\;;\\
-{48xy(x^2+3xy+y^2)\over(x-y)^7}\Theta(y,x)
-{2(3x^3+61x^2y+61xy^2+3y^3)\over(x-y)^6}& \\
+{4(x^4+3x^3y-45x^2y^2-25xy^3)\ln x-4(y^4+3y^3x-45x^2y^2-25yx^3)\ln y\over(x-y)^7}
\;,&x<y\;.\end{array}\right.
\label{varphi4}
\end{eqnarray}

Here, the function $\Theta(x,y)$ is defined as
\begin{eqnarray}
&&\Theta(x,y)=\ln x\ln{y\over x}-2\ln(x-y)\ln{y\over x}-2Li_2({y\over x})+{\pi^2\over3}\;.
\label{theta}
\end{eqnarray}

\end{document}